%% file: m9103.tex
\documentclass{ecai} 

\usepackage{latexsym}
\usepackage{amssymb}
\usepackage{amsmath}
\usepackage{amsthm}
\usepackage{booktabs}
\usepackage{enumitem}
\usepackage{graphicx}
\usepackage{color}

\usepackage{multirow}
\usepackage{graphicx}
\usepackage{cleveref}
\usepackage{amssymb}

\newcommand{\BibTeX}{B\kern-.05em{\sc i\kern-.025em b}\kern-.08em\TeX}

\begin{document}


\begin{frontmatter}


\paperid{9103} 


\title{IPBA: Imperceptible Perturbation Backdoor Attack in Federated Self-Supervised Learning}


\author[A]{\fnms{Jiayao}~\snm{Wang}}
\author[A]{\fnms{Yang}~\snm{Song}}
\author[B]{\fnms{Zhendong}~\snm{Zhao}}
\author[A]{\fnms{Jiale}~\snm{Zhang}}
\author[A]{\fnms{Junwu}~\snm{Zhu}\thanks{Corresponding Author. Email: jwzhu@yzu.edu.cn}}
\author[C]{\fnms{Qilin}~\snm{Wu}} 
\author[D]{\fnms{Dongfang}~\snm{Zhao}} 

\address[A]{School of Information Engineering, Yangzhou University, China}
\address[B]{Institute of Information Engineering, Chinese Academy of Sciences, China}
\address[C]{School of Computing and Artificial Intelligence, Chaohu University, China}
\address[D]{Tacoma School of Engineering and Technology, University of Washington, USA}


\begin{abstract}
Federated self-supervised learning (FSSL) combines the advantages of decentralized modeling and unlabeled representation learning, serving as a cutting-edge paradigm with strong potential for scalability and privacy preservation.
Although FSSL has garnered increasing attention, research indicates that it remains vulnerable to backdoor attacks.
Existing methods generally rely on visually obvious triggers, which makes it difficult to meet the requirements for stealth and practicality in real-world deployment.
In this paper, we propose an imperceptible and effective backdoor attack method against FSSL, called IPBA.
Our empirical study reveals that existing imperceptible triggers face a series of challenges in FSSL, particularly limited transferability, feature entanglement with augmented samples, and out-of-distribution properties. 
These issues collectively undermine the effectiveness and stealthiness of traditional backdoor attacks in FSSL.
To overcome these challenges, IPBA decouples the feature distributions of backdoor and augmented samples, and introduces Sliced-Wasserstein distance to mitigate the out-of-distribution properties of backdoor samples, thereby optimizing the trigger generation process.
Our experimental results on several FSSL scenarios and datasets show that IPBA significantly outperforms existing backdoor attack methods in performance and exhibits strong robustness under various defense mechanisms.
\end{abstract}
\end{frontmatter}

\maketitle
\input{ecai-template/sec/1_introduction}
\input{ecai-template/sec/2_related_work}
\input{ecai-template/sec/3_motivation}
\input{ecai-template/sec/4_methodology}

\input{ecai-template/sec/5_evaluation}

\input{ecai-template/sec/6_conclusion}



\begin{ack}
This work was supported by grants 24KJB520042 (Jiangsu), 2025YSZ-017 (Yangzhou), 2023SGJ014 (Hefei), COGOS-2023HE01 (iFLYTEK), Y202352288 (Zhejiang), and 2023AY11057 (Jiaxing), as well as by resources from Microsoft Azure and the NSF-supported Chameleon testbed.
\end{ack}



\bibliography{mybibfile}

\newpage
\input{ecai-template/sec/Appendix}

\end{document}

%% file: ecai-template/sec/1_introduction.tex
\section{Introduction}

In recent years, self-supervised learning (SSL)~\cite{Moco,Simclr,swav,BYOL} has emerged as a powerful paradigm in machine learning, particularly in computer vision. 
The main advantage of SSL lies in its ability to learn rich representations from large amounts of unlabeled data, bypassing the labor-intensive and costly manual labeling process.
SSL in computer vision aims to develop image encoders that produce similar embeddings for similar images. 
To achieve this, similar image pairs are typically constructed by applying various augmentations to the same image.
The pre-trained encoder can then be used to train downstream classifiers for various tasks. 
These downstream classifiers generally use compact networks with fewer parameters, improving training efficiency and reducing computational cost.

\begin{figure}[t]
  \centering
  \includegraphics[width=1\linewidth]{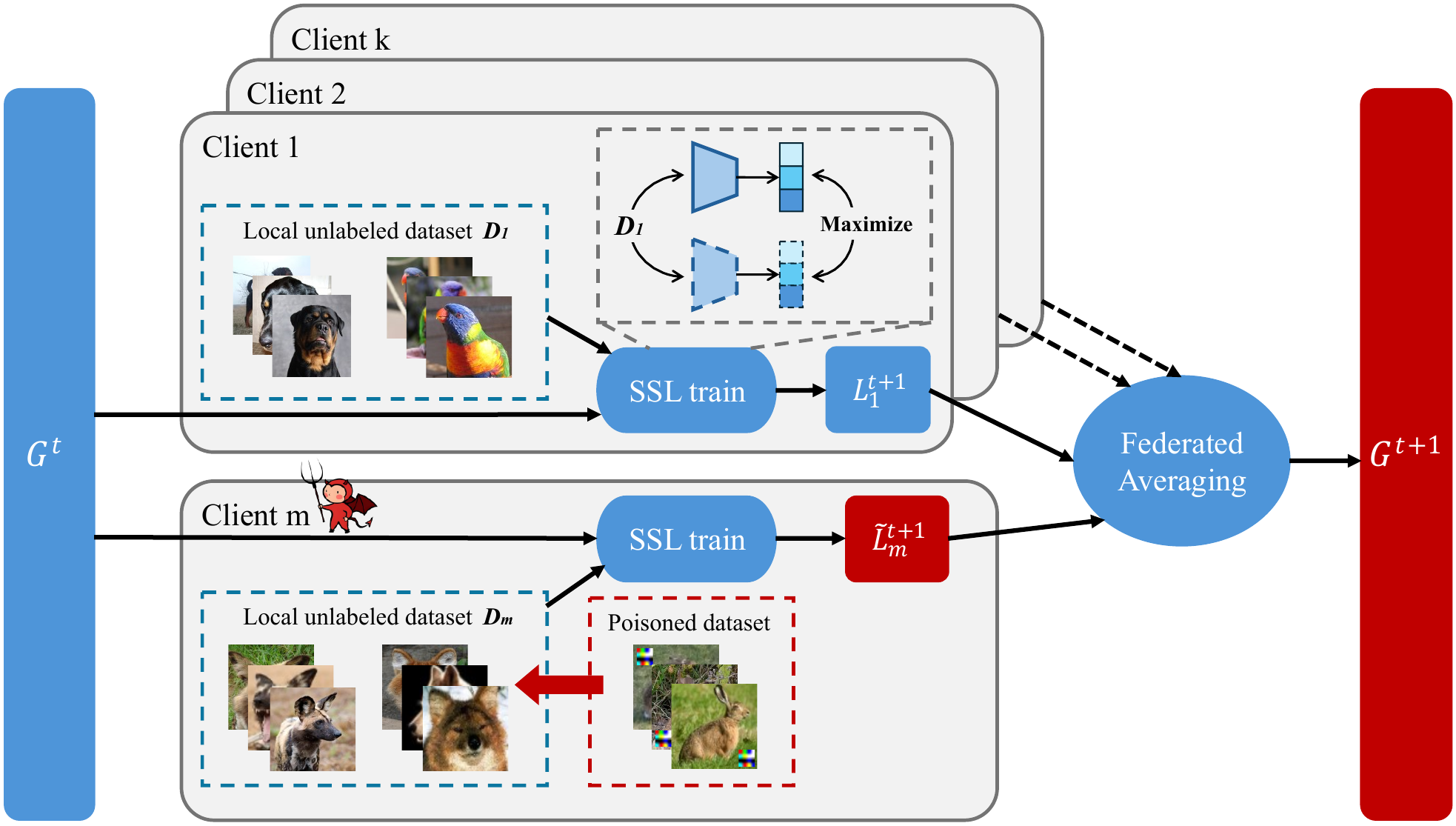}
   \caption{Backdoor injection process in federated self-supervised learning.}
   \label{fig:one}
\end{figure}

Although SSL has made significant progress with unlabeled data, it requires vast amounts of data to achieve performance comparable to supervised learning. This substantial data requirement can be overwhelming for individuals or organizations and may become a significant barrier to model training in data-scarce domains.
In this context, federated self-supervised learning (FSSL)~\cite{FedEMA,L-dawa} employs a distributed learning approach, providing an effective alternative to the data-sharing problem.
Modern edge devices have access to vast amounts of data suitable for training models, giving FSSL a significant advantage. 
However, the decentralized nature of traditional federated supervised learning (FSL) makes it vulnerable to backdoor attacks~\cite{howto,IBA}, and this vulnerability may be inherited in FSSL, posing a potential security risk.

Recent studies~\cite{BADFSS,EmInspector,uba} have shown that FSSL is vulnerable to backdoor attacks. 
Figure \ref{fig:one} illustrates the backdoor injection process in FSSL. 
During this process, malicious clients inject backdoors into the global model by training their local models with poisoned samples that include triggers. 
Specifically, benign clients \( i \in \{1, 2, \dots, N\} \) use local unlabeled datasets \( D_i \) for self-supervised learning, and the Federated Averaging (FedAvg) mechanism aggregates their updated local model \( L_i^{t+1} \) to the central server to update the global model \( G^{t+1} \). 
Meanwhile, malicious clients (e.g., client \( m \)) train on poisoned data, generating backdoored models \( \tilde{L}_m^{t+1} \), which are also uploaded to the server. 
Through the FedAvg, all client models (including those from malicious clients with backdoors) are aggregated into the global model \( G^{t+1} \), embedding the backdoor into the updated global model.
When the global encoder is used to construct downstream tasks, the backdoored classifier predicts each input embedded with the attacker's selected trigger as the corresponding target class specified by the attacker.

However, a common drawback of these existing attack methods is that their trigger patterns are highly visible, making them vulnerable to detection through manual inspection or defense mechanisms.
In this paper, we propose a backdoor attack in FSSL that is both effective and stealthy.

A feasible approach is to apply existing imperceptible triggers~\cite{Wanet,IBA,CTRL} for use in the FSSL setting.
However, experimental results indicate that these imperceptible triggers, which were originally designed for supervised learning (SL), exhibit limited effectiveness when applied to self-supervised models.
Our observations suggest that the primary reason for this ineffectiveness lies in the feature distribution entanglement between backdoor samples and the augmented samples used in FSSL.
Specifically, the transformations induced by the backdoor triggers share similarities with the image augmentations inherent to contrastive learning in FSSL (such as RandomGrayscale and ColorJitter).
As a result, local SSL models struggle to distinguish between the feature distributions of backdoor samples and augmented samples.

Building on the above observations, we propose an imperceptible perturbation backdoor attack that decouples the feature distributions of backdoor samples and augmented samples. 
Specifically, this approach involves increasing the distributional gap between the backdoor samples and augmented samples during the local SSL process, thereby enhancing their separability.
Additionally, to further ensure the stealthiness of the trigger, IPBA imposes a distance constraint on the backdoor samples in the feature space using Sliced Wasserstein Distance~\cite{SWD}, effectively reducing the out-of-distribution properties of the backdoor samples.

Our main contributions are summarized as follows:
\begin{itemize}[itemsep=0pt, parsep=0pt, topsep=0pt]
    \item \textbf{We propose an imperceptible perturbation backdoor attack method:} In FSSL, we propose an innovative backdoor attack method, IPBA. By decoupling the feature distributions of backdoor samples and augmented samples, this method significantly enhances both the stealthiness and effectiveness of backdoor attacks.

    \item \textbf{We propose to apply Sliced Wasserstein Distance:} To mitigate the out-of-distribution properties of backdoor samples in the feature space, we innovatively introduce Sliced Wasserstein Distance. This approach effectively reduces the outlier phenomenon of backdoor samples by minimizing the distance between backdoor and clean samples in the feature space.

    \item \textbf{Extensive experimental evaluation:} We conduct comprehensive evaluations of the proposed method on five public benchmark datasets (CIFAR10, STL10, GTSRB, SVHN, and Tiny-ImageNet). The experimental results show that our method significantly outperforms existing backdoor attack methods in terms of performance and demonstrates strong generalization across various settings. Additionally, we further explore potential defense strategies against IPBA and find that current state-of-the-art defense methods have limitations, emphasizing the urgent need for tailored defense mechanisms.
\end{itemize}

%% file: ecai-template/sec/2_related_work.tex
\section{Related Work}

\subsection{Federated Self-Supervised Learning}
Federated self-supervised learning has gained increasing attention due to its ability to jointly learn representations across decentralized clients without relying on labeled data, while preserving data privacy. 
Early studies~\cite{early2} explored the direct integration of classical self-supervised learning (SSL) methods such as SimCLR and BYOL into the federated setting. 
To address data heterogeneity, approaches like SSFL~\cite{SSFL} and FedEMA~\cite{FedEMA} introduced personalization and momentum-based model updates, respectively. 
Other efforts, such as FedCA~\cite{FedCA} and L-DAWA~\cite{L-dawa}, focused on enhancing model aggregation via global dictionary learning or divergence-aware weighting. 
However, existing research primarily focuses on performance optimization and modeling data heterogeneity, while overlooking potential backdoor risks.

\subsection{Backdoor Attacks}
Attackers conducting backdoor attacks typically select a stealthy trigger and embed it into a subset of training samples to poison the training data.
Traditional backdoor techniques rely on explicit labels~\cite{IBA}, manipulating sample labels to guide the model toward learning attacker-specified behaviors.
However, the absence of labeled data in SSL renders traditional backdoor paradigms inapplicable, thereby prompting the emergence of novel methodologies for backdooring SSL.
BASSL~\cite{BASSL} embeds triggers into target-class images and leverages cropping-based augmentations to generate diverse poisoned views.
BadEncoder~\cite{badencoder} fine-tunes a pre-trained encoder using a trigger-injected shadow dataset to precisely steer model behavior.
Additionally, CorruptEncoder \cite{CorruptEncoder} and PoisonedEncoder \cite{PoisonedEncoder} respectively propose poisoning strategies that target the victim’s training dataset.
Recent studies have demonstrated that FSSL is also susceptible to backdoor threats~\cite{BADFSS,uba,EmInspector}. 
For example, BADFSS~\cite{BADFSS} injects backdoor triggers into the global encoder by leveraging supervised contrastive learning and attention alignment.
However, existing methods typically rely on visually perceptible triggers, making them easily detectable by humans or automated detection systems.
In contrast, our approach surpasses existing methods in both stealth and effectiveness.

\subsection{Backdoor Defenses}
Backdoor attacks have attracted considerable attention due to their high stealthiness and potential for severe damage, which makes it a challenge to effectively defend against such attacks during model training.
On the one hand, reverse engineering-based methods (e.g., Neural Cleanse~\cite{Neural_cleanse} and DECREE~\cite{DECREE}) aim to reconstruct the trigger from a backdoored model and identify its corresponding target class.
Successful trigger reconstruction typically indicates that the encoder is compromised. 
On the other hand, sample-level detection methods identify anomalies by analyzing the influence of input samples on model predictions.
For example, STRIP~\cite{STRIP} measures prediction consistency under input perturbations to detect potential poisoned samples, while GradCAM~\cite{Grad-cam} utilizes activation map visualization to localize trigger regions and assist in identifying backdoored inputs.
We use the above state-of-the-art defense techniques to evaluate our newly proposed attack.

%% file: ecai-template/sec/3_motivation.tex
\section{Observations and Intuitions}

\begin{figure*}[h]
  \centering
  \includegraphics[width=1\linewidth]{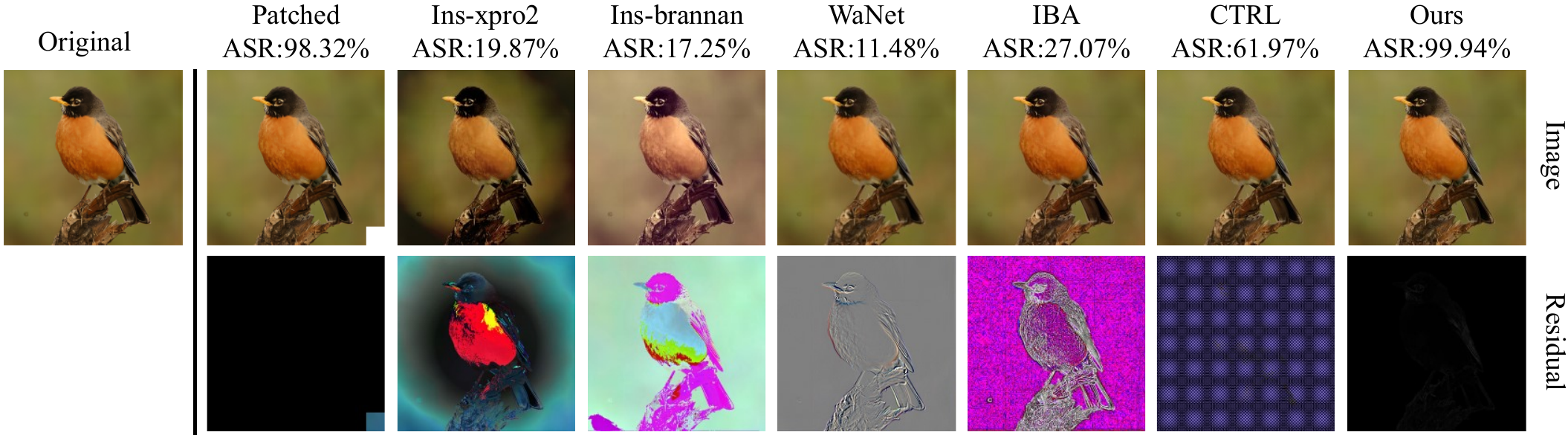}
   \caption{Comparison of clean, backdoored samples created by Patch trigger, Instagram filter trigger, WaNet trigger, IBA trigger,  CTRL trigger and ours. The ASR under the BADFSS threat model is shown next to each method name. Residuals are the difference between clean and backdoored images.}
   \label{fig:two}
\end{figure*}

\textbf{Observation I: Limited transferability.} Existing imperceptible triggers are not as effective as expected in the FSSL environment.
As shown in Figure \ref{fig:two}, we present the attack success rate (ASR) of existing imperceptible triggers on the FSSL model.
We transferred the comparative methods to the FSSL scenario, using the attack methods introduced in BADFSS~\cite{BADFSS} and replacing the original patch triggers with imperceptible triggers.
The results show that although these imperceptible triggers achieve high ASR in SL and SSL, they do not perform well in the FSSL environment.\\
\textbf{Observation II: Feature entanglement.} The augmented samples are entangled with the backdoor samples in the feature distribution.
Specifically, we use a pre-trained ResNet18~\cite{resnet} for binary classification to determine whether IBA-injected backdoor samples and contrastively augmented samples are distinguishable.
Experimental results indicate that the model struggles to differentiate between the two types of samples.
The t-SNE visualization in the left panel of \Cref{fig:three} shows significant overlap between the two types of samples in the feature space.
Based on this observation, we infer that the reduced effectiveness of supervised backdoor attacks in FSSL may be attributed to the high similarity in feature distribution between contrastively augmented samples and backdoor samples, which weakens the model’s ability to distinguish between the two.
In contrast, our method (right panel of \Cref{fig:three}) effectively decouples the feature distributions of poisoned and augmented samples in the latent space.\\
\textbf{Observation III: Out-of-distribution property.} Existing backdoor samples exhibit out-of-distribution properties.
Tao et al.~\cite{DRUPE} present a critical insight that current SSL attacks introduce strong backdoor signals into the embedding space, pushing malicious samples outside the clean data distribution.
Inspired by this insight, we conducted experiments to verify whether a similar phenomenon occurs in FSSL.
Specifically, we fed both clean pre-training data and poisoned samples into the same encoder to extract their feature embeddings. 
Then, Principal Component Analysis (PCA) was applied to reduce the dimensionality of these embeddings for visualization in a two-dimensional space. 
As shown in the left panel of \Cref{fig:four}, existing FSSL attacks also exhibit such out-of-distribution properties.\\
\textbf{Intuition and Design Motivation.}
Based on the above observations (more results are provided in the Appendix) which highlight challenges such as limited transferability, feature entanglement, and out-of-distribution properties, we identify several key factors that are essential for achieving stealthy backdoor attacks in FSSL:
\begin{itemize}[itemsep=0pt, parsep=0pt, topsep=0pt]
    \item Decoupling the feature distributions of backdoor samples and augmented samples during local client pre-training. Since the augmentation strategies in the pre-training stage are known in advance, we design \(\mathcal{L}_{dis}\) to quantify the distributional gap between the two batches of images.

    \item Introducing a dual alignment loss tailored for malicious clients. To further enhance the effectiveness of the backdoor attack, we design \(\mathcal{L}_{align}\) to pull close the features of backdoor images and target images.

    \item An excessive distributional gap may reduce the visual naturalness of backdoor images and compromise the stealthiness of the trigger. Therefore, we design \(\mathcal{L}_{ste}\) to seamlessly fuse the backdoor with the original image and eliminate out-of-distribution properties in the feature space.
\end{itemize}

%% file: ecai-template/sec/4_methodology.tex
\section{Methodology}

In this section, we first introduce preliminaries of FSSL system model, Wasserstein Distance, and threat model. We then elaborate on the design of the Poisoned Data Constructor phase. Finally, we describe the Backdoor Injection strategy and formulate its corresponding optimization objective.

\subsection{Threat Model} 
\textbf{Attack Objective.} 
In FSSL, the attacker's goal is to inject a backdoor into the global model while maintaining both effectiveness and stealthiness.
Effectiveness means the model consistently misclassifies trigger inputs into a target class, yielding high ASR.
Stealthiness means the backdoor does not noticeably degrade the model’s main-task performance, thus evading detection.
Overall, the attacker aims to induce the global SSL model to exhibit malicious behavior on specific inputs, without significantly affecting its overall utility or being exposed during training.\\
\textbf{Attack Knowledge and Capabilities.} 
We assume the attacker masquerades as a benign FL participant, with knowledge of the global model and full control of local training.
This includes the ability to manipulate the local dataset (e.g., by embedding triggers), redesign the loss function, and arbitrarily alter the model training procedure.
Therefore, the attacker also has access to the data augmentation strategies used for pretraining the encoder, which can be leveraged in IPBA to generate stealthy and effective backdoor triggers.
However, the attacker is generally unaware of other critical details of the FSSL system, such as the models from benign clients and the aggregation rules employed by the server.

\begin{figure}[t]
  \centering
  \includegraphics[width=0.9\linewidth]{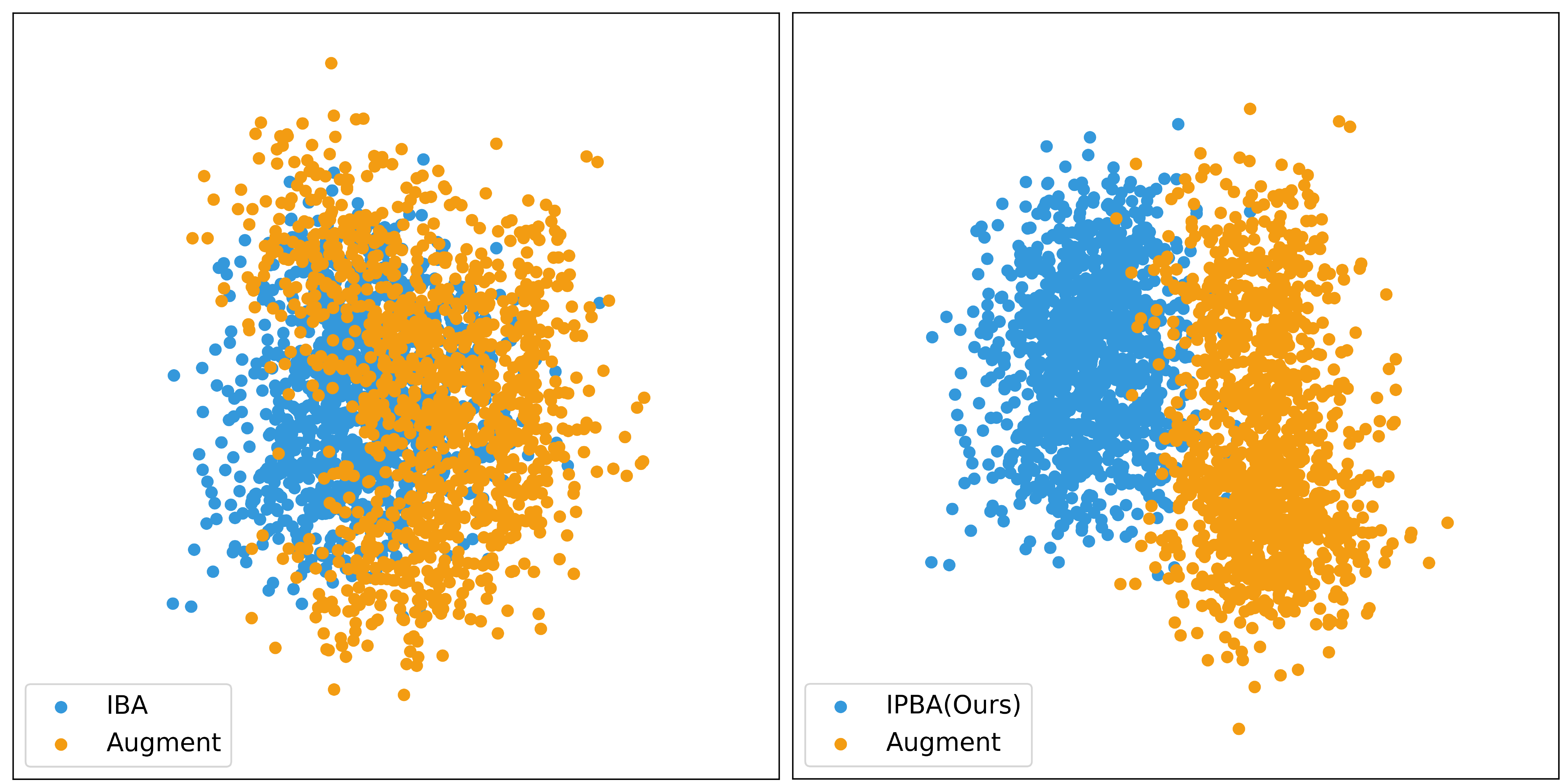}
   \caption{The t-SNE visualization of feature vectors in the latent space under different attacks.}
   \label{fig:three}
\end{figure}
\begin{figure}[t]
  \centering
  \includegraphics[width=0.9\linewidth]{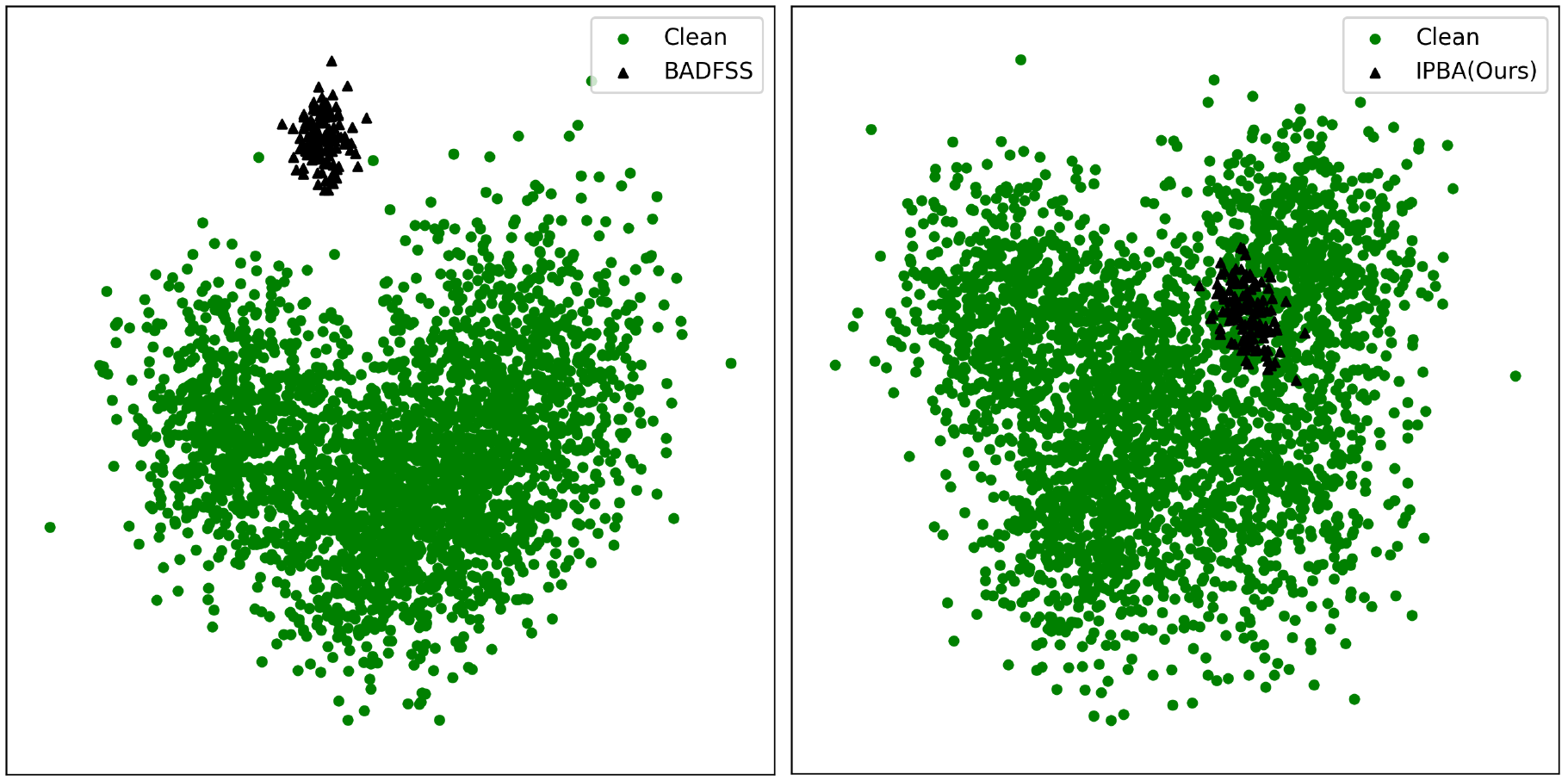}
   \caption{PCA visualization of clean and poisoned sample embeddings in backdoored models under different attacks.}
   \label{fig:four}
\end{figure}
\subsection{Preliminaries}

\textbf{FSSL System Model.}
We consider a standard FSSL framework, in which a total of \(N\) decentralized clients collaboratively train a global encoder \(\mathbf{W}_{global}\). In each communication round, every selected client updates its local model using its private unlabeled dataset \(\mathcal{D}_i = \{\mathbf{x}_1, \dots, \mathbf{x}_{|\mathcal{D}_i|}\}\), where \(i \in \mathcal{N} = \{1, 2, \dots, N\}\). The training proceeds iteratively through the following three steps:\\
\textbf{Step 1: Global Model Distribution.} At the beginning of round \(t\), the central server broadcasts the current global encoder \(\mathbf{W}^t_{global}\) to a subset of selected clients \(\mathcal{S}^t\).\\
\textbf{Step 2: Local Self-Supervised Training.} Each client \(i \in \mathcal{S}^t\) initializes its local model with the received global encoder \(\mathbf{W}^t_{global}\), and performs \(e\) local training epochs on its unlabeled dataset \(\mathcal{D}_i\). The resulting updated local encoder is denoted by \(\mathbf{W}_i^{t}\), which is then uploaded to the central server.\\
\textbf{Step 3: Federated Aggregation.} Upon receiving all local updates, the server aggregates them to obtain a new global encoder \(\mathbf{W}^{t+1}_{global}\). Following the FedAVG protocol, the aggregation is performed as:
\(\mathbf{W}^{t+1}_{global} = \sum_{i \in \mathcal{S}^t} p_i^t \mathbf{W}_i^{t},\)
where the aggregation weight is given by \(p_i^t = \frac{|\mathcal{D}_i|}{\sum_{j \in \mathcal{S}^t} |\mathcal{D}_j|}.\)

After finishing the training process, we freeze the parameters of the global model, and use it to construct a downstream predictor with only a small amount of labeled data.\\
\textbf{Wasserstein Distance.}
Wasserstein distance (WD) is a widely used metric for measuring the discrepancy between underlying distributions, particularly in scenarios where the common support or density functions are unknown.
As our goal is to minimize the distributional difference between poisoned and clean samples, Wasserstein distance offers a smooth and effective solution to this problem.
We adopt the 2-Wasserstein distance, also known as the Earth Mover’s Distance, which is defined as:
\begin{equation}
\mathcal{W}(\zeta, \tau) = \left( \inf_{\psi \in \Pi(\zeta, \tau)} \int_{(u, v) \sim \psi} p(u, v) | u - v |_2 , du , dv \right)^{1/2},
\label{eq:one}
\end{equation}
where \(\zeta\) and \(\tau\) represent the marginal probability distributions of clean and poisoned samples in the feature space, estimated from their corresponding empirical embeddings.  \(\psi\)  denotes the joint distribution between the two.
The infimum \(\inf\) denotes the greatest lower bound of the computed distance on the joint distribution. 
The integral term calculates the sum of the distance between every two data points \((u, v)\) drawn from the joint distribution \(\psi\). \(p(u, v)\) is the probability of jointly drawing the two samples. \(||u - v||_2\) is the \(L^2\) distance between the two samples.

\subsection{Poisoned Data Constructor}
In the FSSL framework, when a malicious client is selected to participate in a training round, it gains the opportunity to influence the global model update. 
Specifically, malicious client \(k\) receives the current global parameters \(\mathbf{W}^t_{global}\), updates them using its local dataset \(\mathcal{D}_k\), and obtains a locally refined model \(\mathbf{W}_k^{t}\). This local model is subsequently incorporated into the next-round global model through the FedAvg aggregation scheme.
Based on this process, we introduce an initial phase termed the Poisoned Data Constructor, in which a trainable backdoor injector \(\mathcal{I}_\psi\) is designed to transform a clean input \(x\) into a poisoned sample \(x' = \mathcal{I}_\psi(x)\).
This sample remains imperceptibly different from the original, but it effectively misleads the model into making incorrect classifications for the target class.
To this end, we design three essential loss functions that jointly optimize the poisoned samples in terms of effectiveness, disentanglement, and stealthiness.\\
\textbf{Dual alignment loss.}
Given that FSSL models generate feature embeddings rather than relying on explicit labels, a malicious client must optimize trigger-injected inputs to align with the target semantics in order to achieve attack effectiveness.
Moreover, as a critical step toward enhancing attack efficiency, the attacker must ensure that its local model can accurately recognize the target semantics. 
We refer to this process as dual feature alignment between the backdoor sample and the target sample.
Following \cite{badencoder}, the dual feature alignment of the malicious client can be formally expressed as:
\begin{align}
\mathcal{L}_{align} &= - \frac{1}{|\mathcal{D}_k|} \sum_{x \in \mathcal{D}_k} \left[ 
s\left( f(x', \tilde{\theta}), f(x_t, \tilde{\theta}) \right) \right. \notag \\
&\quad \left. - s\left( f(x_t, \tilde{\theta}), f(x_t, \theta) \right)
\right],
\label{eq:two}
\end{align}
where \(s(\cdot, \cdot)\) is the cosine similarity function, \(f(\cdot, \theta)\) denotes the feature encoder, \(x'\) is the poisoned sample generated by the injector \(\mathcal{I}_\psi\), \(x_t\) is the target class sample, and \(\tilde{\theta}, \theta\) are the backdoored and clean model parameters, respectively.\\
\textbf{Feature disentanglement loss.}
To mitigate the severe overlap between poisoned samples and their augmented counterparts in the feature space, we design a feature disentanglement loss, inspired by~\cite{IMPERATIVE}. 
This loss maximizes the representational distance between the two, encouraging the encoder to learn more separable feature embeddings in the latent space, thereby enhancing the recognizability and robustness of the backdoor semantics.
Specifically, we first convert the images from the RGB color space to HSV and HSL, as these color representations better capture the variations introduced by augmentation operations such as ColorJitter, which randomly perturbs brightness, contrast, saturation, and hue. 
Then, we compute the differences between the poisoned images and their augmented counterparts in both color spaces, based on which the following loss function is defined:
\begin{equation}
\mathcal{L}_{dis} = \mathbb{E}_{x \sim \mathcal{D}_k} \left[ 
\sum_{c \in \{H, S, V, L\}} \left\| c(x') - c(\tilde{x}) \right\|_2^2 
\right],
\label{eq:three}
\end{equation}
where \(H, S, V, L\) represent the Hue, Saturation, Value, and Lightness channel transformations from the HSV and HSL color spaces, respectively. 
The input sample \(x\) augmented by the transformations used during the encoder's pre-training stage is denoted as \(\tilde{x}\).
The function \( c(\cdot) \) denotes the color channel transformation for a specific channel (\(H, S, V, L\)), and \( \|u-v\|_2^2 \) denotes the squared \( \ell_2 \) distance between sample \( u \) and sample \( v \).\\
\textbf{Stealthiness loss.}
To address the out-of-distribution characteristics of poisoned samples, we introduce a stealthiness loss aimed at minimizing the distributional discrepancy between poisoned and clean samples. 
The 2-Wasserstein distance serves as an effective metric for this purpose. However, directly computing WD in high-dimensional spaces is challenging due to the involved optimization process.
To overcome this, we adopt the Sliced Wasserstein Distance (SWD)~\cite{SWD}, a variant that projects high-dimensional data onto multiple random one-dimensional subspaces, computes WD in each, and averages the results.
This approach significantly improves computational efficiency and numerical stability, as the 1D 2-Wasserstein distance admits a closed-form solution.
\begin{equation}
\mathcal{W}_{sliced}(\zeta, \tau) = \left( 
\frac{1}{S} \sum_{s=1}^{S} \int_{0}^{1} 
\left\| F_c^s(z) - F_b^s(z) \right\|_2 \, dz 
\right)^{\frac{1}{2}},
\label{eq:four}
\end{equation}
where \( S \) is the number of one-dimensional directions. \( F^s_c \) and \( F^s_b \) represent the projections of the clean and poisoned embeddings into one-dimensional data points along the direction of slice \( s \), respectively.

We first extract high-dimensional feature representations of poisoned and clean samples using a pre-trained model \(F\), and then employ the SWD to effectively measure the distributional discrepancy between them. 
The final stealthiness loss is defined as follows:
\begin{equation}
\mathcal{L}_{ste} = \sum_{x \in \mathcal{D}_k} \text{SWD}\left(F(x'), F(x)\right).
\label{eq:five}
\end{equation}

Based on the above loss formulations and design objectives, we define the optimization of the backdoor injector \( \mathcal{I}_\psi \) as a joint learning task:
\begin{equation}
\text{arg} \min_{\mathcal{I}_\psi} \; \mathcal{L}_{injector} = \mathcal{L}_{ste} + \alpha \cdot \mathcal{L}_{dis} + \beta \cdot \mathcal{L}_{align},
\label{eq:six}
\end{equation}
where \(\alpha\) and \(\beta\) are hyperparameters used to balance the contributions of the three loss terms.

\subsection{Backdoor Injection}
The core objective of the second-stage backdoor injection is to guide the model to learn backdoor behaviors through poisoned data during local training and progressively migrate these behaviors into the global model.
Specifically, each selected malicious client is required to embed backdoor features into its local model while maintaining stealthiness. 
To achieve this, the attacker utilizes trigger-embedded samples to construct feature representations that are semantically similar to the target class, thereby inducing the model to produce the desired responses to backdoor inputs.
In addition, to enhance the effectiveness and efficiency of the attack, the attacker guides the backdoored model to produce feature representations for reference data that are similar to those generated by the clean model, thereby ensuring accurate recognition of target-class samples by the backdoor model (as shown in \Cref{eq:two}).\\
\textbf{Utility loss.}
To ensure that the injection process does not significantly degrade the performance on the main task, the attacker must maintain the local model’s ability to represent clean samples correctly, such that its output features remain consistent with those of the clean model. 
Based on this, we define the utility loss \(\mathcal{L}_{uti}\) as follows:
\begin{equation}
\mathcal{L}_{uti} = -\frac{1}{|\mathcal{D}_k|} \sum_{x \in \mathcal{D}_k} s\left( f(x, \tilde{\theta}), f(x, \theta) \right).
\label{eq:even}
\end{equation}
Combining \(\mathcal{L}_{align}\) and \(\mathcal{L}_{uti}\), the objective of local encoder backdoor injection can be formulated as the following optimization problem:
\begin{equation}
\arg\min_{\tilde{\theta}} \mathcal{L}_{encoder} = \lambda_1 \cdot \mathcal{L}_{align}(\tilde{\theta}) + \lambda_2 \cdot \mathcal{L}_{uti}(\tilde{\theta}),
\label{eq:eight}
\end{equation}
where \(\lambda_1\) and \(\lambda_2\) are hyperparameters that balance the two loss components.

%% file: ecai-template/sec/5_evaluation.tex
\section{Evaluation}

To demonstrate the effectiveness and stealthiness of our approach, we implemented IPBA using Pytorch and compared its performance with existing state-of-the-art backdoor attack methods. 
All experiments were conducted on an NVIDIA 4090 GPU. We designed comprehensive experiments to address the following three research questions:\\
\textbf{RQ1 (Effectiveness of IPBA):} Can IPBA successfully inject backdoors into FSSL?\\
\textbf{RQ2 (Stealthiness of IPBA):} Can IPBA achieve good stealthiness and naturalness across different evaluation metrics?\\
\textbf{RQ3 (Robustness of IPBA):} Can IPBA effectively resist existing defense methods?
\begin{table*}[t]
  \caption{Camparison of attack performance on different datasets. The best result are \textbf{highlighted}.}
  \centering
  \resizebox{\textwidth}{!}{%
  \begin{tabular}{@{}c|c|ccccccccccc@{}}
    \toprule
    \textbf{Pre-training} & \textbf{Downstream} & \textbf{Benign} & \multicolumn{2}{c}{\textbf{WaNet}~\cite{Wanet}} & \multicolumn{2}{c}{\textbf{IBA}~\cite{IBA}} & \multicolumn{2}{c}{\textbf{CTRL}~\cite{CTRL}} & \multicolumn{2}{c}{\textbf{BADFSS}~\cite{BADFSS}} & \multicolumn{2}{c}{\textbf{Ours}} \\
    \cmidrule(l){3-13} 
    \textbf{Dataset} & \textbf{Dataset} & CA & BA$\uparrow$ & ASR$\uparrow$ & BA$\uparrow$ & ASR$\uparrow$ & BA$\uparrow$ & ASR$\uparrow$ & BA$\uparrow$ & ASR$\uparrow$ & BA$\uparrow$ & ASR$\uparrow$ \\
    \midrule
    \multirow{3}{*}{CIFAR-10} & STL-10 & 75.14 & 73.27 & 11.91 & 73.21 & 10.14 & 75.73 & 66.85 & 72.36 & 66.35 & 72.82 & \textbf{96.11} \\
                              & GTSRB  & 82.84 & 77.16 & 7.71  & 77.36 & 35.41 & 77.53 & 62.07 & 75.32 & 70.21 & 78.64 & \textbf{93.68} \\
                              & SVHN   & 63.52 & 56.86 & 12.45 & 58.67 & 30.91 & 60.35 & 45.91 & 65.64 & 56.38 & 71.03 & \textbf{92.83} \\
    \midrule
    \multirow{3}{*}{STL-10}   & CIFAR-10 & 85.21 & 82.06 & 11.48 & 84.77 & 27.07 & 78.19 & 61.97 & 77.52 & 68.32 & 87.19 & \textbf{99.94} \\
                              & GTSRB    & 76.32 & 79.84 & 4.41  & 81.65 & 17.29 & 70.37 & 61.28 & 72.32 & 69.93 & 74.83 & \textbf{97.41} \\
                              & SVHN     & 56.47 & 55.74 & 15.24 & 55.07 & 14.13 & 53.77 & 49.36 & 54.97 & 54.66 & 60.42 & \textbf{99.53} \\
    \midrule
    \multirow{3}{*}{Tiny-ImageNet} & STL-10 & 89.58 & 87.14 & 12.60 & 87.51 & 10.44 & 82.26 & 47.14 & 75.71 & 51.29 & 87.15 & \textbf{99.91} \\
                                   & GTSRB  & 78.32 & 77.26 & 10.51 & 80.72 & 9.38  & 70.91 & 49.85 & 68.13 & 46.54 & 75.75 & \textbf{96.91} \\
                                   & SVHN   & 73.67 & 72.63 & 13.93 & 71.94 & 18.75 & 65.94 & 40.91 & 60.38 & 41.97 & 71.86 & \textbf{95.25} \\
    \bottomrule
  \end{tabular}
  }
  \label{tab:one}
\end{table*}
\subsection{Experimental Setup}

\textbf{Datasets.} 
Five datasets are employed in the experiments including CIFAR-10~\cite{cifar10}, STL-10~\cite{stl10}, GTSRB~\cite{GTSRB}, SVHN~\cite{shvn}, and Tiny-ImageNet~\cite{imagenet}. More details about the used datasets can be found in Appendix.\\
\textbf{Evaluation Metrics.}  
Similar to existing work in~\cite{badencoder,DRUPE}, we evaluated the effectiveness of all attack methods using three metrics: Clean Accuracy (CA), Attack Success Rate (ASR), and Backdoored Accuracy (BA). 
A well-executed backdoor attack should maximize the ASR while maintaining a high BA. More details are explained in Appendix.
To assess the stealthiness and naturalness of our IPBA, we used three metrics: SSIM~\cite{ssim}, PSNR~\cite{psnr}, and LPIPS~\cite{lpips}.
In the experiments, higher SSIM and PSNR values, along with lower LPIPS, indicate better stealthiness and naturalness of the generated backdoored images.\\
\textbf{Baseline.}
We compare IPBA with the state-of-the-art backdoor attack method, BADFSS~\cite{BADFSS}, and use WaNet~\cite{Wanet}, IBA~\cite{IBA}, and CTRL~\cite{CTRL} as baseline triggers.
These methods significantly outperform earlier backdoor attack methods in terms of stealthiness.
Following the experimental setup in~\cite{BADFSS}, we adapt all baseline methods to the FSSL scenario for evaluation and strictly follow the original implementations.\\
\textbf{Implementation Details.} 
We use SimCLR as the default self-supervised learning algorithm and employ ResNet-18~\cite{resnet} as the default architecture network for the encoders.
Moreover, we use a two-layer multi-layer perceptron (MLP) as a predictor.
Following previous work~\cite{BADFSS,badencoder,IMPERATIVE}, we  set the decay rate $m$ = 0.99, batch size $B$ = 256, SGD as optimizer with learning rate $lr$ = 0.001 and run experiments with $K$ = 5 clients (one is malicious and the poison ratio is 1\%) for $E$ = 200 training rounds, where each client performs $e$ = 3 local epochs in each round.
We use the U-Net architecture~\cite{u-net} for the backdoor injector.

\subsection{Effectiveness Evaluation (RQ1)}

\begin{figure}[t]
  \centering
  \includegraphics[width=1\linewidth]{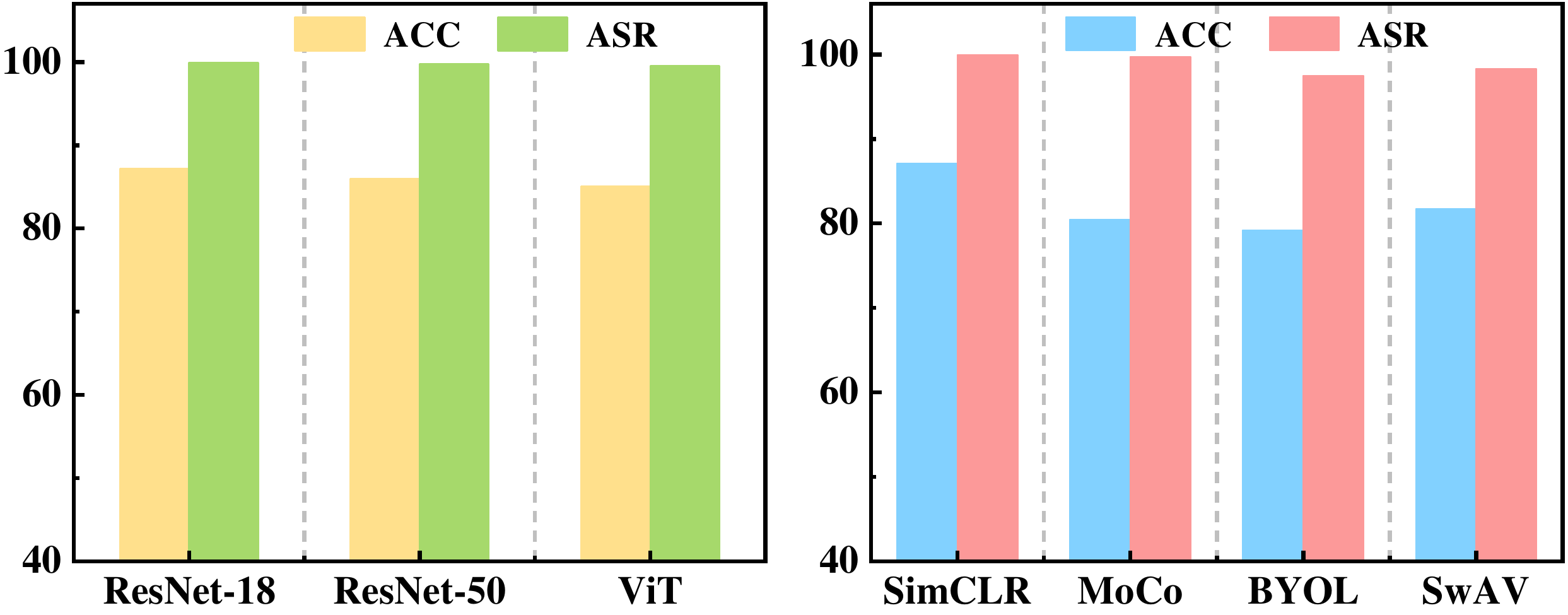}
   \caption{Experimental results for different encoder architectures and SSL algorithms.}
   \label{fig:five}
\end{figure}

\textbf{Effectiveness comparison with SOTA attack methods.} 
To evaluate the effectiveness of IPBA, we compared its ASR and BA with four SOTA attack methods. 
The experiments followed a standard SSL setup, where the pre-training and downstream datasets were different. 
\Cref{tab:one} shows the performance of different attack methods. The results indicate that IPBA achieves a high ASR while maintaining a high BA. 
Specifically, with STL-10 as the pre-training dataset and CIFAR-10 as the downstream dataset, IPBA achieved the best ASR (99.94\%) and BA (87.19\%). 
Compared to supervised-based attack methods (e.g., WaNet and IBA), IPBA outperforms them in terms of ASR across different downstream datasets, which validates our previous conclusion that supervised-based methods are unsuitable for the FSSL scenario. 
Furthermore, compared to self-supervised-based attack methods (e.g., CTRL and BADFSS), IPBA also shows superior performance in terms of ASR and BA across all datasets.\\
\textbf{Effectiveness on different SSL algorithms.} 
An effective attack method should exhibit strong generalizability and be adaptable to various SSL algorithms. 
To this end, we evaluate the performance of IPBA in FSSL under four representative SSL methods: SimCLR~\cite{Simclr}, MoCo~\cite{Moco}, BYOL~\cite{BYOL}, and SwAV~\cite{BYOL}. 
\Cref{fig:five} presents the attack performance of IPBA across these SSL algorithms.
The results demonstrate that IPBA exhibits stable attack performance across different SSL algorithms, highlighting its adaptability and generalizability.\\
\textbf{Effectiveness on different encoder architectures.}
To evaluate the effectiveness of IPBA across different encoder architectures, we conducted experiments on STL-10 using three representative architectures: ResNet-18~\cite{resnet}, ResNet-50~\cite{resnet}, and ViT~\cite{Vit}.
\Cref{fig:five} illustrates the attack performance of IPBA on these encoder architectures.
The results show that IPBA can successfully inject backdoors into various encoder architectures while maintaining high BA classification performance, highlighting its strong generalizability.
\begin{table}[t]
  \caption{Stealthness evaluation on different datasets.}
  \centering
  \resizebox{\columnwidth}{!}{%
  \begin{tabular}{@{}c|c|ccc@{}}
    \toprule
    \textbf{Pre-training} & \textbf{Downstream} & \multirow{2}{*}{\textbf{SSIM}$\uparrow$} & \multirow{2}{*}{\textbf{PSNR}$\uparrow$} & \multirow{2}{*}{\textbf{LPIPS}$\downarrow$} \\
    \textbf{Dataset} & \textbf{Dataset} \\
    \midrule
    \multirow{3}{*}{CIFAR-10} & STL-10 & 0.9142 & 22.19 & 0.0311 \\
                              & GTSRB  & 0.9914 & 35.03 & 0.0017 \\
                              & SVHN   & 0.9946 & 35.22 & 0.0012 \\
    \midrule
    \multirow{3}{*}{STL-10}   & CIFAR-10 & 0.9898 & 35.38 & 0.0026 \\
                              & GTSRB    & 0.9911 & 36.88 & 0.0041 \\
                              & SVHN     & 0.9857 & 31.89 & 0.0032 \\
    \midrule
    \multirow{3}{*}{Tiny-ImageNet} & STL-10 & 0.9889 & 33.46 & 0.0044 \\
                                  & GTSRB  & 0.9908 & 32.68 & 0.0031 \\
                                  & SVHN   & 0.9846 & 31.57 & 0.0041 \\                              
    \bottomrule
  \end{tabular}
  }
  \label{tab:two}
\end{table}
\subsection{Stealthiness Evaluation (RQ2)}
To evaluate the stealthiness of IPBA, we compared the trigger images generated by IPBA with those of SOTA attack methods.
Additionally, we used PCA to visualize the embedding spaces of clean and poisoned samples on the backdoored models.\\
\textbf{Stealthiness results from the perspective of images.} 
Figure 2 compares the poisoned samples generated by IPBA with those from SOTA attack methods, along with their magnified residuals (×2).
The results show that the residual generated by IPBA is the smallest, leaving only a few subtle artifacts.

Moreover, we evaluated the stealthiness of the triggers generated by IPBA using three metrics: PSNR, SSIM, and LPIPS.
\Cref{tab:two} presents the evaluation results across different datasets.
As shown in the table, the SSIM values are consistently close to 1, with most exceeding 0.98, indicating that the structural changes introduced by IPBA are minimal.
The PSNR values are notably high, suggesting that the noise introduced is nearly imperceptible.
Additionally, the LPIPS values are extremely low, ranging from 0.0012 to 0.0311, further confirming that the perceptual difference between the original and perturbed images is negligible.
Overall, these results demonstrate that IPBA effectively maintains the stealthiness of the images, introducing virtually no noticeable visual changes.\\
\textbf{Stealthiness results from the perspective of latent space.} 
Many backdoor defense methods assume that poisoned and benign samples are clearly separated in the latent space. 
Therefore, ensuring the stealthiness of the attack method from the latent space perspective is crucial.
We visualized the embedding features of BADFSS using PCA, as shown in Figure 4.
In the left diagram, we observe that the features of BADFSS form two distinct clusters, indicating that they can be easily detected in the latent space by any clustering algorithm.
In contrast, in our IPBA, the feature representations of poisoned samples are intermingled with those of benign samples, forming a single cluster.
This demonstrates that IPBA achieves optimal stealthiness in the latent space, breaking the assumption of latent separation, and effectively evading backdoor defenses.
\begin{table}[t]
  \caption{Defense evaluation results.}
  \centering
  \resizebox{\columnwidth}{!}{%
  \begin{tabular}{@{}c|c|cc@{}}
    \toprule
    \textbf{Pre-training} & \textbf{Downstream} & {\textbf{Neural Cleanse~\cite{Neural_cleanse}}} & {\textbf{DECREE~\cite{DECREE}}}\\
    \cmidrule(l){3-4}
    \textbf{Dataset} & \textbf{Dataset} & Anomaly Index & $\mathcal{PL}^1$-Norm\\
    \midrule
    \multirow{3}{*}{CIFAR-10} & STL-10 & 1.11 & 0.26 \\
                              & GTSRB  & 1.16 & 0.37 \\
                              & SVHN  & 1.32 & 0.19  \\
    \midrule
    \multirow{3}{*}{STL-10}  & CIFAR-10 & 0.98 & 0.23 \\
                             & GTSRB    & 1.22 & 0.31 \\
                             & SVHN    & 1.37 & 0.25 \\
    \bottomrule
  \end{tabular}
  }
  \label{tab:three}
\end{table}
\subsection{Robustness Evaluation (RQ3)}
To evaluate the robustness of IPBA against existing backdoor defenses, we implemented representative backdoor defense methods (i.e., DECREE~\cite{DECREE}, STRIP~\cite{STRIP}, Neural Cleanse~\cite{Neural_cleanse}, and Grad-CAM~\cite{Grad-cam}) and assessed IPBA's resistance to these defenses. Additional defense results are provided in the appendix.\\
\textbf{Resistance to DECREE.} 
DECREE~\cite{DECREE} detects backdoor attacks in pre-trained encoders by reversing the trigger. 
If the reversed trigger has a $\mathcal{PL}^1$-Norm smaller than 0.1, the encoder is considered compromised.
As shown in Table 3, all backdoor triggers generated by IPBA have a $\mathcal{PL}^1$-Norm above 0.1, successfully evading detection by DECREE.\\
\textbf{Resistance to STRIP.}
STRIP~\cite{STRIP} is a sample-based backdoor detection method. 
It assumes that a backdoored model’s predictions exhibit stability on malicious samples, detecting such samples by computing entropy after overlaying random samples.
Figure 6 shows that STRIP fails to establish an effective threshold to distinguish benign from malicious samples, allowing our attack to bypass detection.\\
\begin{figure}[t]
  \centering
  \includegraphics[width=1\linewidth]{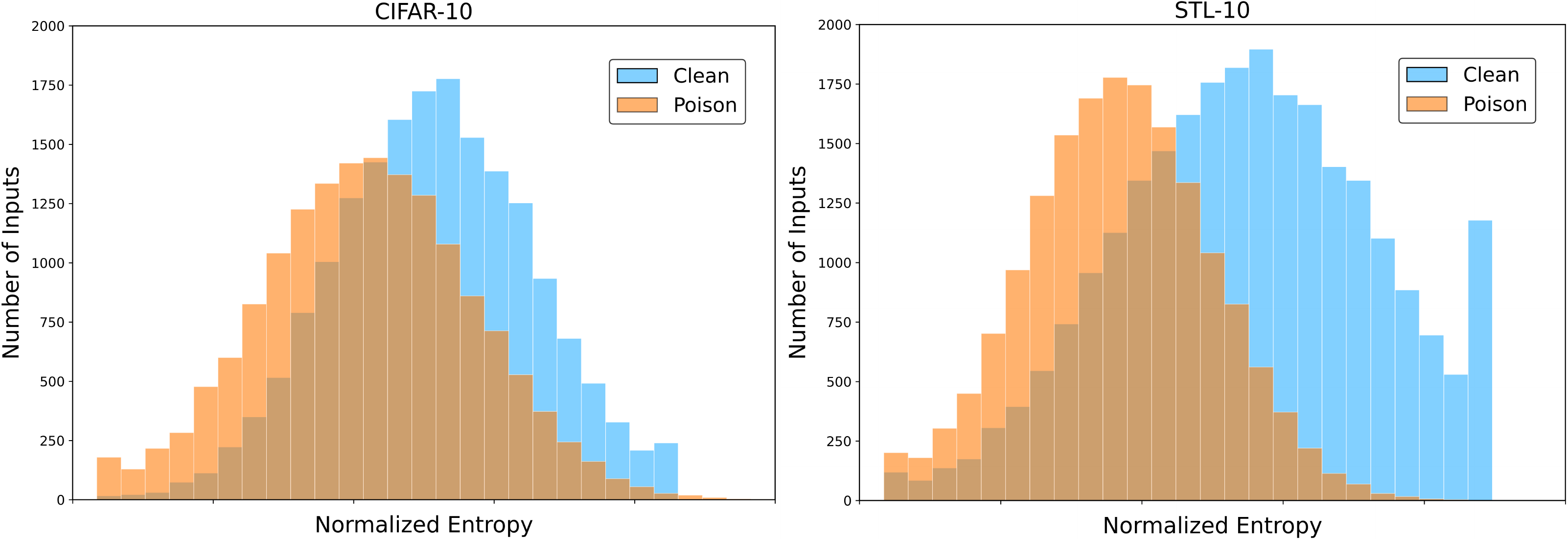}
   \caption{Experimental results of STRIP.}
   \label{fig:six}
\end{figure}
\begin{figure}[t]
  \centering
  \includegraphics[width=1\linewidth]{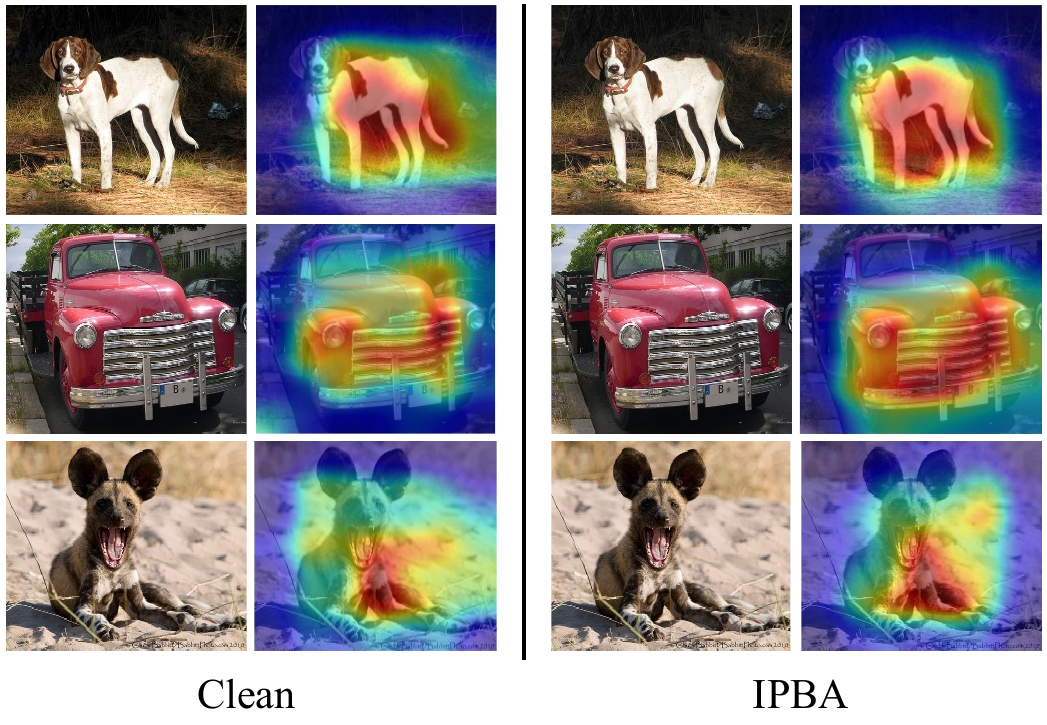}
   \caption{GradCAM visualization results for both clean and backdoored models.}
   \label{fig:seven}
\end{figure}
\textbf{Resistance to Neural Cleanse.} 
Neural Cleanse (NC)~\cite{Neural_cleanse} is a defense method against trigger generation that measures the deviation of the reconstructed trigger by calculating the anomaly index, marking models with an anomaly index greater than 2 as backdoored.
Since NC is designed specifically for classifiers and cannot be directly applied to image encoders, we use NC to identify backdoors in downstream classifiers.
Experimental results (see Table 3) show that the anomaly index of IPBA is below 2 across all datasets, successfully evading detection by NC.\\
\textbf{Resistance to GradCAM.} 
GradCAM~\cite{Grad-cam} generates heatmaps to show the contribution of each pixel in the model's prediction, with backdoored models typically exhibiting abnormal heatmaps.
Figure 7 shows the visualization heatmaps of a clean model and a backdoored model attacked by IPBA. 
From this figure, we can find that the heatmaps of these models are very similar, and IPBA is capable of resisting defense methods based on GradCAM.

\subsection{Ablation Study}
In this section, we demonstrate the effects of each important component in Section 4.3 by ablating them respectively.
As shown in \Cref{tab:four}, after ablating each component, the ASR of our method decreases across the target downstream datasets.
Specifically, when ablating $\mathcal{L}_{dis}$ and $\mathcal{L}_{align}$ sequentially, the ASR of IPBA decreases from 99.94\% to 42.32\% and 9.68\%, respectively, which is consistent with our previous observations.
When ablating $\mathcal{L}_{ste}$, the SSIM of IPBA decreases from 0.9898 to 0.1937, the PSNR drops from 35.38 to 6.79, and the LPIPS increases from 0.0026 to 0.6534. These changes clearly indicate that the absence of key components significantly weakens the effectiveness of our backdoor attack method.

\begin{table}[t]
    \caption{Performance of Ablation Studies.}
    \centering
    \resizebox{\columnwidth}{!}{%
    \begin{tabular}{c c c| c c c c c}
        \toprule
        \textbf{$\mathcal{L}_{ste}$} & $\mathcal{L}_{dis}$ & $\mathcal{L}_{align}$ & \textbf{BA}$\uparrow$ & \textbf{ASR}$\uparrow$ & \textbf{SSIM}$\uparrow$ & \textbf{PSNR}$\uparrow$ & \textbf{LPIPS}$\downarrow$\\
        \midrule
        $\checkmark$ & $\checkmark$ & $\checkmark$ & 87.13 & 99.94 & 0.9898 & 35.38 & 0.0026 \\
        \midrule
        $\checkmark$ &  & $\checkmark$ & 90.35 & 42.32 & 0.9864 & 32.08 & 0.0048 \\
        \midrule
        $\checkmark$ & $\checkmark$ &  & 62.25 & 9.68 & 0.9802 & 30.16 & 0.0046 \\
        \midrule
         & $\checkmark$ & $\checkmark$ & 84.66 & 95.27 & 0.1937 & 6.79 & 0.6534 \\
        \bottomrule
    \end{tabular}
    }
    \label{tab:four}
\end{table}

%% file: ecai-template/sec/6_conclusion.tex
\section{Conclusion}

This paper introduces IPBA, an imperceptible and effective backdoor attack method for FSSL. 
IPBA decouples the feature distributions of backdoor and augmented samples and incorporates Sliced Wasserstein Distance to effectively address the challenges and limitations of existing imperceptible triggers in FSSL.
Results show that IPBA outperforms baseline methods and is effective under different settings.
We further explore potential countermeasures against our attack and find that existing defense mechanisms are insufficient to mitigate IPBA, indicating the need for specifically designed defense strategies to alleviate backdoor attacks in FSSL.

%% file: ecai-template/sec/Appendix.tex
\appendix

\section{Appendix}

\subsection{Convergence Preservation under IPBA}

We provide a theoretical analysis showing that the proposed Imperceptible Perturbation Backdoor Attack (IPBA) does not hinder the convergence of federated self-supervised learning (FSSL), provided that the utility-preserving constraint is enforced during attack generation.

\subsubsection*{Setup}
Let $f(x; \theta): \mathcal{X} \to \mathbb{R}^d$ be the feature encoder parameterized by $\theta$. Let $\mathcal{L}_{\text{SSL}}^{(i)}(\theta)$ be the local self-supervised contrastive loss used by client $i$ in round $t$.

Assume that the global model $\theta_t$ is updated via FedAvg over $N$ clients:
\[
\theta_{t+1} = \theta_t - \eta \sum_{i=1}^N p_i \, g_t^{(i)}, \quad \text{where } p_i = \frac{n_i}{\sum_{j=1}^N n_j},
\]
and $g_t^{(i)}$ is the (local) gradient estimate returned by client $i$.

Suppose client $k$ is malicious, and instead of minimizing $\mathcal{L}_{\text{SSL}}^{(k)}$, it minimizes a modified loss:
\[
\mathcal{L}_{\text{IPBA}}^{(k)}(\theta) := \lambda_1 \mathcal{L}_{\text{align}}(\theta) + \lambda_2 \mathcal{L}_{\text{uti}}(\theta),
\]
where:
\begin{itemize}
    \item $\mathcal{L}_{\text{align}}(\theta)$ encourages trigger samples to align with target class representations.
    \item $\mathcal{L}_{\text{uti}}(\theta)$ penalizes deviation from the clean model in representation space:
    \[
    \mathcal{L}_{\text{uti}}(\theta) = \mathbb{E}_{x \sim \mathcal{D}_k} \| f(I_{\psi}(x); \theta) - f(x; \theta_0) \|^2,
    \]
    where $I_{\psi}(x)$ is the backdoored input and $\theta_0$ is the clean reference model.
\end{itemize}

Let $\nabla \mathcal{L}_{\text{SSL}}^{(i)}(\theta)$ denote the gradient under clean training, and $g_t^{(i)}$ be the actual gradient sent by client $i$ at round $t$. For honest clients, $g_t^{(i)} = \nabla \mathcal{L}_{\text{SSL}}^{(i)}(\theta_t)$. For the malicious client $k$, let
\[
\| g_t^{(k)} - \nabla \mathcal{L}_{\text{SSL}}^{(k)}(\theta_t) \| \leq \epsilon
\]
for some small $\epsilon > 0$ due to the utility constraint.

\subsubsection*{Assumptions}
\begin{itemize}
    \item[A1.] Each $\mathcal{L}_{\text{SSL}}^{(i)}$ is $L$-smooth: $\| \nabla \mathcal{L}_{\text{SSL}}^{(i)}(\theta) - \nabla \mathcal{L}_{\text{SSL}}^{(i)}(\theta') \| \leq L \| \theta - \theta' \|$.
    \item[A2.] The norm of local gradients is bounded: $\| \nabla \mathcal{L}_{\text{SSL}}^{(i)}(\theta) \| \leq G$ for all $i$.
    \item[A3.] The learning rate $\eta$ satisfies $\eta \leq 1/L$.
\end{itemize}

\subsubsection*{Main Result}
\textbf{Theorem.} Under assumptions A1--A3, suppose there is at most one malicious client $k$ injecting IPBA updates such that $\| g_t^{(k)} - \nabla \mathcal{L}_{\text{SSL}}^{(k)}(\theta_t) \| \leq \epsilon$ for all $t$. Then after $T$ rounds of federated learning, the global model satisfies:
\[
\min_{0 \leq t \leq T-1} \mathbb{E}[\| \nabla \mathcal{L}_{\text{SSL}}(\theta_t) \|^2] \leq \frac{2(\mathcal{L}_{\text{SSL}}(\theta_0) - \mathcal{L}_{\text{SSL}}^*)}{\eta T} + O(\rho \epsilon^2),
\]
where $\rho = p_k$ is the weight of the malicious client, and $\mathcal{L}_{\text{SSL}} = \sum_i p_i \mathcal{L}_{\text{SSL}}^{(i)}$.

\begin{proof}
From $L$-smoothness of $\mathcal{L}_{\text{SSL}}$, we expand:
\[
\mathcal{L}_{\text{SSL}}(\theta_{t+1}) \leq \mathcal{L}_{\text{SSL}}(\theta_t) - \eta \langle \nabla \mathcal{L}_{\text{SSL}}(\theta_t), g_t \rangle + \frac{L \eta^2}{2} \| g_t \|^2,
\]
where $g_t = \sum_i p_i g_t^{(i)}$ is the weighted average gradient (including the malicious one).

Decompose $g_t$ into clean and malicious parts:
\[
\begin{aligned}
  g_t &= \sum_{i \neq k} p_i \nabla \mathcal{L}_{\text{SSL}}^{(i)}(\theta_t) + p_k g_t^{(k)} \\
      &= \nabla \mathcal{L}_{\text{SSL}}(\theta_t) + p_k (g_t^{(k)} - \nabla \mathcal{L}_{\text{SSL}}^{(k)}(\theta_t)).
\end{aligned}
\]

Then:
\[
\begin{aligned}
\langle \nabla \mathcal{L}_{\text{SSL}}(\theta_t), g_t \rangle &= \| \nabla \mathcal{L}_{\text{SSL}}(\theta_t) \|^2 + p_k \langle \nabla \mathcal{L}_{\text{SSL}}(\theta_t), g_t^{(k)} - \nabla \mathcal{L}_{\text{SSL}}^{(k)}(\theta_t) \rangle \\
&\geq \| \nabla \mathcal{L}_{\text{SSL}}(\theta_t) \|^2 - p_k \epsilon \| \nabla \mathcal{L}_{\text{SSL}}(\theta_t) \|.
\end{aligned}
\]

Also, since $\| g_t \| \leq G$ from A2, we get:
\[
\mathcal{L}_{\text{SSL}}(\theta_{t+1}) \leq \mathcal{L}_{\text{SSL}}(\theta_t) - \eta \| \nabla \mathcal{L}_{\text{SSL}}(\theta_t) \|^2 + \eta p_k \epsilon \| \nabla \mathcal{L}_{\text{SSL}}(\theta_t) \| + \frac{L \eta^2 G^2}{2}.
\]

Using inequality $ab \leq \frac{a^2}{2} + \frac{b^2}{2}$ on the middle term, we obtain:
\[
\mathcal{L}_{\text{SSL}}(\theta_{t+1}) \leq \mathcal{L}_{\text{SSL}}(\theta_t) - \frac{\eta}{2} \| \nabla \mathcal{L}_{\text{SSL}}(\theta_t) \|^2 + \frac{\eta p_k^2 \epsilon^2}{2} + \frac{L \eta^2 G^2}{2}.
\]

Summing over $t=0$ to $T-1$ and rearranging yields the stated result.
\end{proof}

\subsection{More observations}
\begin{figure}[t]
  \centering
  \includegraphics[width=1\linewidth]{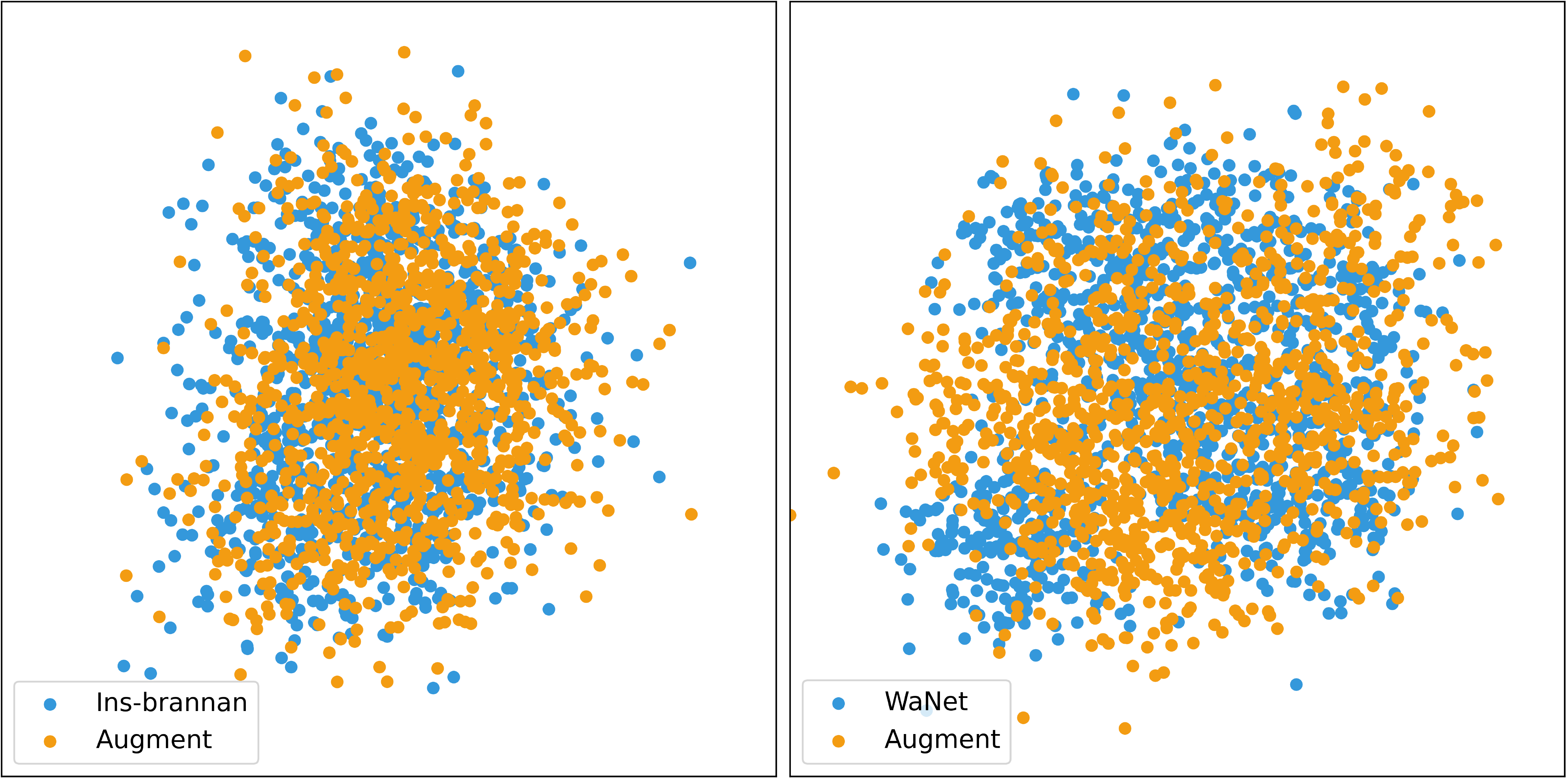}
   \caption{The t-SNE Visualization of feature vectors in the latent space under different attacks.}
   \label{fig:eight}
\end{figure}

\textbf{Observation II: Feature entanglement.} We provide additional experimental results to support our observation on feature entanglement. 
\Cref{fig:eight} shows the feature distribution of augmented and backdoor samples across two different methods: "Ins-brannan" (left panel) and "WaNet" (right panel).

\begin{figure}[t]
  \centering
  \includegraphics[width=1\linewidth]{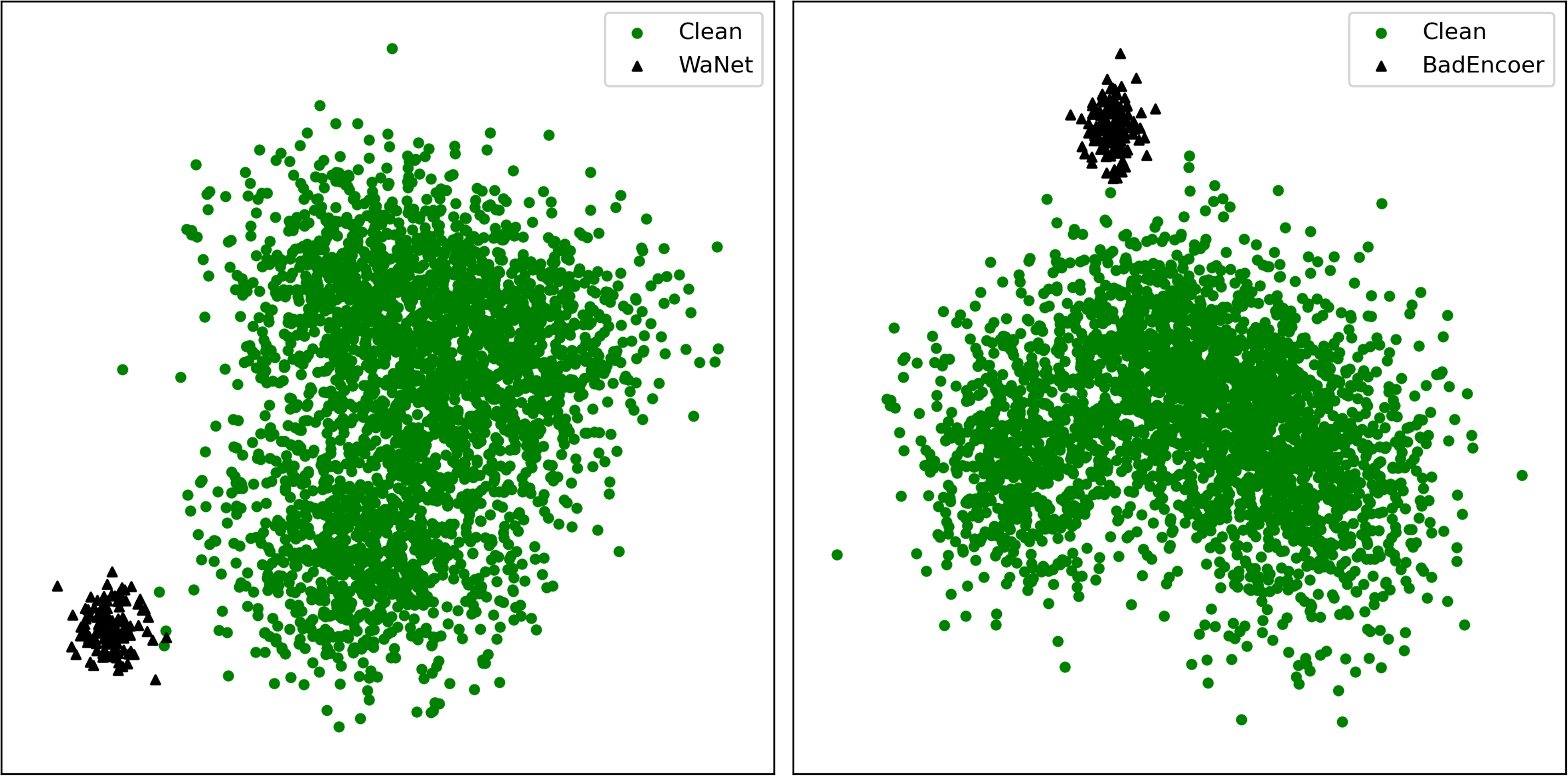}
   \caption{PCA visualization of clean and poisoned sample embeddings in backdoored models under different attacks.}
   \label{fig:nine}
\end{figure}

In both cases, we observe that the augmented samples (represented by orange dots) are heavily entangled with the backdoor samples (represented by blue dots), forming overlapping clusters in the feature space. This overlap further corroborates our previous finding that the augmented and backdoor samples exhibit high similarity in feature distribution, making it difficult for the model to effectively distinguish between them. These results emphasize the impact of feature entanglement on the performance of backdoor attacks in the FSSL scenario.\\
\textbf{Observation III: Out-of-distribution property.} We provide additional experimental results to support our third observation regarding the out-of-distribution properties of backdoor samples. 
\Cref{fig:nine} shows the feature distribution of clean samples (green) and backdoor samples generated by two different methods: "WaNet" (left panel) and "BadEncoder" (right panel).

In both cases, we observe that the backdoor samples (black triangles) are positioned far from the clean samples (green circles) in the feature space, indicating out-of-distribution behavior. This further corroborates our previous findings that backdoor samples injected using these methods exhibit significant deviations from the clean data distribution, making them easily detectable when analyzed in the feature space. These results highlight the out-of-distribution nature of the backdoor samples, which can impact the effectiveness of certain defense methods.

\subsection{Dataset Details}
We use the following datasets in our method evaluation.

\begin{itemize}
    \item \textbf{CIFAR10}: This dataset comprises 60,000 images of $32 \times 32 \times 3$ pixels and 10 different classes for basic image recognition tasks, divided into 50,000 for training and 10,000 for testing.
    \item \textbf{STL10}: STL10 includes 5,000 labeled training images and 8,000 for testing with a resolution of $96 \times 96 \times 3$ pixels, across 10 classes. Additionally, it provides 100,000 unlabeled images for unsupervised learning. Notably, they are resized to $32 \times 32 \times 3$ to be consistent with other datasets.
    \item \textbf{GTSRB}: This dataset includes 51,800 images of traffic signs categorized into 43 classes. It is split into 39,200 training and 12,600 testing images, each sized at $32 \times 32 \times 3$.
    \item \textbf{SVHN}: SVHN is a dataset of digit images from house numbers in Google Street View, consisting of 73,257 training and 26,032 testing images, each $32 \times 32 \times 3$ in size.
    \item \textbf{Tiny-ImageNet}: Tiny-ImageNet is designed for large-scale object classification, featuring 100,000 training samples and 10,000 testing samples across 200 categories. Each image has a resolution of $224 \times 224$ pixels with three color channels.
\end{itemize}

\subsection{Evaluation Metrics Details}
The meanings of evaluation metrics are shown below:
\begin{itemize}
    \item \textbf{CA:} This is the accuracy of a clean downstream classifier in correctly classifying the clean testing images from the corresponding downstream dataset.
    \item \textbf{BA:} BA measures the accuracy of a backdoored downstream classifier on the same clean testing images from the corresponding downstream dataset.
    \item \textbf{ASR:} ASR measures the percentage of these backdoored test images that are classified as the target class by the backdoored downstream classifier.
\end{itemize}

\begin{table}[t]
    \caption{Performance of IPBA under different FL defense methods.}
    \centering
    \resizebox{\columnwidth}{!}{%
    \begin{tabular}{c c c c c }
        \toprule
        {Defense} & \multicolumn{2}{c}{CIFAR-10} & \multicolumn{2}{c}{STL-10}\\
         \cmidrule(lr){2-3}\cmidrule(l){4-5}
         Method& BA$\uparrow$ & ASR$\uparrow$ & BA$\uparrow$ & ASR$\uparrow$\\
        \midrule
        Krum & 72.37 & 95.67 & 75.15 & 97.29 \\
        \midrule
        Trimmed-Mean & 74.12 & 90.45 & 81.42 & 98.41 \\
        \midrule
        Foolsgold & 69.61 & 92.23 & 78.31 & 97.11 \\
        \midrule                          
        FLAME & 70.23 & 94.46 & 80.74 & 96.91 \\    
        \midrule                          
        FLARE & 71.52 & 93.29 & 79.32 & 99.71 \\ 
        \midrule                          
        EmInspector & 72.13 & 95.47 & 80.25 & 98.32 \\ 
        \bottomrule
    \end{tabular}
    }
    \label{tab:five}
\end{table}
\begin{figure}[htbp]
  \centering
  \includegraphics[width=1\linewidth]{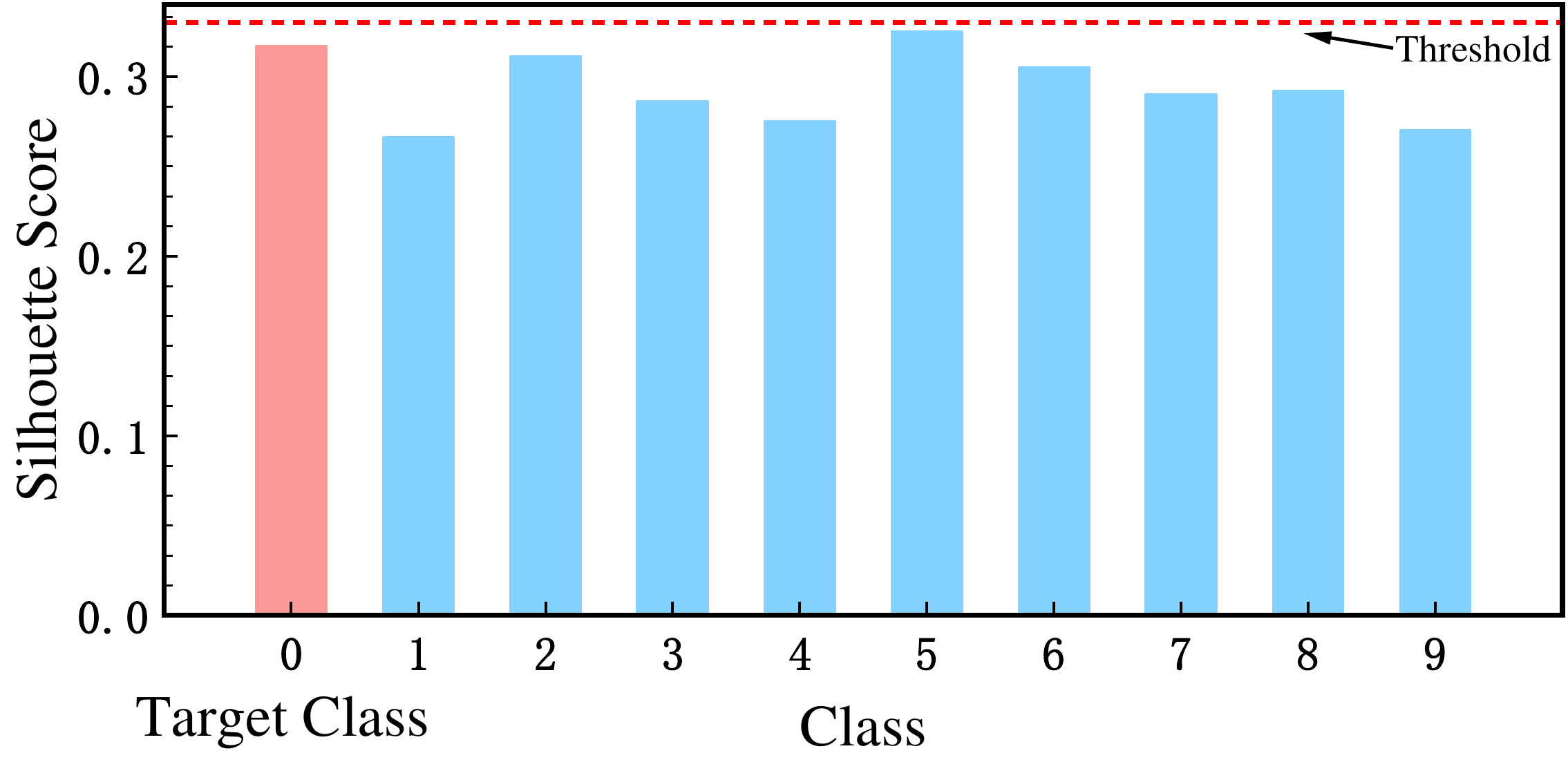}
   \caption{Experimental results of AC.}
   \label{fig:ten}
\end{figure}


\subsection{Supplementary Experimental Results}
\textbf{Resistance to Activation Clustering.} 
Activation Clustering (AC) is a data inspection method designed to detect backdoor attacks. 
It trains the model on potentially poisoned data and collects the activations from the penultimate layer. 
AC assumes that poisoned data in the target class will form a separate cluster that is either small or distant from the class center. 
It identifies potential poisoning by calculating the silhouette score for each class, where a higher score indicates a better fit to two distinct clusters. 
Given the assumption that an attacker cannot poison more than half of the data, the smaller cluster is considered to be the poisoned data.
Assuming data labels are available, we evaluate the effectiveness of AC against our attack method.

\Cref{fig:ten} shows that AC fails to identify the target class (class 0), with its silhouette score even lower than those of non-target classes (e.g., class 5), and is thus unable to effectively detect the poisoned samples.
This may be attributed to our introduced stealthiness loss, which encourages the feature representations of backdoor samples to closely resemble those of clean samples, leading to strong entanglement between the two and significantly weakening the separability of clustering-based methods.\\
\textbf{Resistance to FL Defense Methods.} 
To assess the robustness of IPBA under established federated learning (FL) defense mechanisms, we evaluated its performance against six representative aggregation-based defenses: Krum, Trimmed-Mean, FoolsGold, FLAME, FLARE, and EmInspector.

As shown in \Cref{tab:five}, IPBA consistently achieves high attack success rates (ASR) across both CIFAR-10 and STL-10 datasets, despite the presence of these defenses. For instance, under Krum, IPBA maintains an ASR of 95.67\% on CIFAR-10 and 97.29\% on STL-10. 
When facing more advanced and robust defenses such as FLARE and FLAME, IPBA still achieves ASRs above 93\% on both datasets, reaching as high as 99.71\% on STL-10 with FLARE.
Even against the latest EmInspector defense mechanism, IPBA continues to demonstrate excellent attack performance.

Meanwhile, the BA of the global models remains relatively stable, indicating that IPBA does not significantly degrade model utility. These results demonstrate that IPBA exhibits strong resilience against a wide range of FL defense methods, highlighting its practicality and stealth in real-world federated settings.\\
\textbf{Effectiveness under varying data distributions.} 
Different non-IID data distributions are critical and realistic considerations in FL scenarios, as clients often possess data with varying underlying distributions.
In our setup, we simulate data heterogeneity using Dirichlet distributions with \(\alpha\) = 0.1 and \(\alpha\) = 10. As shown in Table 6, IPBA maintains stable performance across these heterogeneous settings.\\
\textbf{Effectiveness on different attack intervals.} 
In specific scenarios, attackers may be randomly chosen in each round, or the number of attacks might be reduced to improve stealthiness.
To this end, we systematically evaluate the performance of IPBA under varying attack intervals.
As shown in Table 7, the results demonstrate that IPBA consistently maintains a high attack success rate across different attack intervals, indicating strong robustness and adaptability.

\begin{table}[t]
    \caption{Performance of IPBA under different data distribution settings.}
    \centering
    \resizebox{\columnwidth}{!}{%
    \begin{tabular}{c| c| c c c c}
        \toprule
        \multirow{2}{*}{Dataset} & \multirow{2}{*}{Setting} & \multicolumn{2}{c}{Clean} & \multicolumn{2}{c}{Backdoored} \\
        \cmidrule(lr){3-6}
         &  & CA$\uparrow$ & ASR$\downarrow$ & BA$\uparrow$ & ASR$\uparrow$ \\
        \midrule
        \multirow{2}{*}{CIFAR-10} & \(\alpha\) = 10 & 75.14 & 10.73 & 72.82 & 96.11 \\
                                 & \(\alpha\) = 0.1     & 67.95 & 5.79 & 66.16 & 90.77 \\
        \midrule
        \multirow{2}{*}{STL-10} & \(\alpha\) = 10 & 85.21 & 12.81 & 87.19 & 99.94 \\
                               & \(\alpha\) = 0.1      & 79.07 & 7.26 & 80.81 & 93.65 \\
        \midrule                       
        \multirow{2}{*}{Tiny-Imagenet} & \(\alpha\) = 10 & 89.58 & 11.72 & 87.15 & 99.91 \\
                                      & \(\alpha\) = 0.1     & 81.24 & 5.28 & 79.16 & 91.42 \\                          
        \bottomrule
    \end{tabular}
    }
    \label{tab:six}
\end{table}

\begin{table}[t]
    \caption{Performance of IPBA under different attack intervals.}
    \centering
    \resizebox{\columnwidth}{!}{%
    \begin{tabular}{c| c c c c c}
        \toprule
        {Attack Interval} & {1} & {2} & {4} & {8} & {16} \\
        \midrule
        {CIFAR-10} & 96.11 & 95.92 & 91.12 & 84.58 & 77.21\\
        \midrule
        {STL-10} & 99.94 & 99.06 & 95.12 & 88.28 & 80.51\\
        \bottomrule
    \end{tabular}
    }
    \label{tab:seven}
\end{table}

%% file: m9103.bbl
\begin{thebibliography}{38}
\providecommand{\natexlab}[1]{#1}
\providecommand{\url}[1]{\texttt{#1}}
\expandafter\ifx\csname urlstyle\endcsname\relax
  \providecommand{\doi}[1]{doi: #1}\else
  \providecommand{\doi}{doi: \begingroup \urlstyle{rm}\Url}\fi

\bibitem[Bagdasaryan et~al.(2020)Bagdasaryan, Veit, Hua, Estrin, and Shmatikov]{howto}
E.~Bagdasaryan, A.~Veit, Y.~Hua, D.~Estrin, and V.~Shmatikov.
\newblock How to backdoor federated learning.
\newblock In \emph{International conference on artificial intelligence and statistics}, pages 2938--2948. PMLR, 2020.

\bibitem[Caron et~al.(2020)Caron, Misra, Mairal, Goyal, Bojanowski, and Joulin]{swav}
M.~Caron, I.~Misra, J.~Mairal, P.~Goyal, P.~Bojanowski, and A.~Joulin.
\newblock Unsupervised learning of visual features by contrasting cluster assignments.
\newblock \emph{Advances in neural information processing systems}, 33:\penalty0 9912--9924, 2020.

\bibitem[Chen et~al.(2020)Chen, Kornblith, Norouzi, and Hinton]{Simclr}
T.~Chen, S.~Kornblith, M.~Norouzi, and G.~Hinton.
\newblock A simple framework for contrastive learning of visual representations.
\newblock In \emph{International conference on machine learning}, pages 1597--1607. PMLR, 2020.

\bibitem[Coates et~al.(2011)Coates, Ng, and Lee]{stl10}
A.~Coates, A.~Ng, and H.~Lee.
\newblock An analysis of single-layer networks in unsupervised feature learning.
\newblock In \emph{Proceedings of the fourteenth international conference on artificial intelligence and statistics}, pages 215--223. JMLR Workshop and Conference Proceedings, 2011.

\bibitem[Dosovitskiy et~al.(2021)Dosovitskiy, Beyer, Kolesnikov, Weissenborn, Zhai, Unterthiner, Dehghani, Minderer, Heigold, Gelly, Uszkoreit, and Houlsby]{Vit}
A.~Dosovitskiy, L.~Beyer, A.~Kolesnikov, D.~Weissenborn, X.~Zhai, T.~Unterthiner, M.~Dehghani, M.~Minderer, G.~Heigold, S.~Gelly, J.~Uszkoreit, and N.~Houlsby.
\newblock An image is worth 16x16 words: Transformers for image recognition at scale.
\newblock In \emph{International Conference on Learning Representations}, 2021.

\bibitem[Feng et~al.(2023)Feng, Tao, Cheng, Shen, Xu, Liu, Zhang, Ma, and Zhang]{DECREE}
S.~Feng, G.~Tao, S.~Cheng, G.~Shen, X.~Xu, Y.~Liu, K.~Zhang, S.~Ma, and X.~Zhang.
\newblock Detecting backdoors in pre-trained encoders.
\newblock In \emph{Proceedings of the IEEE/CVF Conference on Computer Vision and Pattern Recognition}, pages 16352--16362, 2023.

\bibitem[Gao et~al.(2019)Gao, Xu, Wang, Chen, Ranasinghe, and Nepal]{STRIP}
Y.~Gao, C.~Xu, D.~Wang, S.~Chen, D.~C. Ranasinghe, and S.~Nepal.
\newblock {STRIP:} a defence against trojan attacks on deep neural networks.
\newblock In \emph{Proceedings of the 35th Annual Computer Security Applications Conference}, pages 113--125. {ACM}, 2019.

\bibitem[Grill et~al.(2020)Grill, Strub, Altch{'e}, Tallec, Richemond, Buchatskaya, Doersch, Avila~Pires, Guo, Gheshlaghi~Azar, et~al.]{BYOL}
J.-B. Grill, F.~Strub, F.~Altch{'e}, C.~Tallec, P.~Richemond, E.~Buchatskaya, C.~Doersch, B.~Avila~Pires, Z.~Guo, M.~Gheshlaghi~Azar, et~al.
\newblock Bootstrap your own latent-a new approach to self-supervised learning.
\newblock \emph{Advances in neural information processing systems}, 33:\penalty0 21271--21284, 2020.

\bibitem[He et~al.(2021)He, Yang, Mushtaq, Lee, Soltanolkotabi, and Avestimehr]{SSFL}
C.~He, Z.~Yang, E.~Mushtaq, S.~Lee, M.~Soltanolkotabi, and S.~Avestimehr.
\newblock Ssfl: Tackling label deficiency in federated learning via personalized self-supervision.
\newblock \emph{arXiv preprint arXiv:2110.02470}, 2021.

\bibitem[He et~al.(2016)He, Zhang, Ren, and Sun]{resnet}
K.~He, X.~Zhang, S.~Ren, and J.~Sun.
\newblock Deep residual learning for image recognition.
\newblock In \emph{Proceedings of the IEEE conference on computer vision and pattern recognition}, pages 770--778, 2016.

\bibitem[He et~al.(2020)He, Fan, Wu, Xie, and Girshick]{Moco}
K.~He, H.~Fan, Y.~Wu, S.~Xie, and R.~Girshick.
\newblock Momentum contrast for unsupervised visual representation learning.
\newblock In \emph{Proceedings of the IEEE/CVF conference on computer vision and pattern recognition}, pages 9729--9738, 2020.

\bibitem[Huynh-Thu and Ghanbari(2008)]{psnr}
Q.~Huynh-Thu and M.~Ghanbari.
\newblock Scope of validity of psnr in image/video quality assessment.
\newblock \emph{Electronics letters}, 44\penalty0 (13):\penalty0 800--801, 2008.

\bibitem[Jia et~al.(2022)Jia, Liu, and Gong]{badencoder}
J.~Jia, Y.~Liu, and N.~Z. Gong.
\newblock Badencoder: Backdoor attacks to pre-trained encoders in self-supervised learning.
\newblock In \emph{2022 IEEE Symposium on Security and Privacy (SP)}, pages 2043--2059. IEEE, 2022.

\bibitem[Kolouri et~al.(2019)Kolouri, Nadjahi, Simsekli, Badeau, and Rohde]{SWD}
S.~Kolouri, K.~Nadjahi, U.~Simsekli, R.~Badeau, and G.~Rohde.
\newblock Generalized sliced wasserstein distances.
\newblock \emph{Advances in neural information processing systems}, 32, 2019.

\bibitem[Krizhevsky et~al.(2009)Krizhevsky, Hinton, et~al.]{cifar10}
A.~Krizhevsky, G.~Hinton, et~al.
\newblock Learning multiple layers of features from tiny images.
\newblock 2009.

\bibitem[Li et~al.(2023)Li, Pang, Xi, Du, Ji, Yao, and Wang]{CTRL}
C.~Li, R.~Pang, Z.~Xi, T.~Du, S.~Ji, Y.~Yao, and T.~Wang.
\newblock An embarrassingly simple backdoor attack on self-supervised learning.
\newblock In \emph{Proceedings of the IEEE/CVF International Conference on Computer Vision}, pages 4367--4378, 2023.

\bibitem[Liu et~al.(2022)Liu, Jia, and Gong]{PoisonedEncoder}
H.~Liu, J.~Jia, and N.~Z. Gong.
\newblock $\{$PoisonedEncoder$\}$: Poisoning the unlabeled pre-training data in contrastive learning.
\newblock In \emph{31st USENIX Security Symposium (USENIX Security 22)}, pages 3629--3645, 2022.

\bibitem[Netzer et~al.(2011)Netzer, Wang, Coates, Bissacco, Wu, Ng, et~al.]{shvn}
Y.~Netzer, T.~Wang, A.~Coates, A.~Bissacco, B.~Wu, A.~Y. Ng, et~al.
\newblock Reading digits in natural images with unsupervised feature learning.
\newblock In \emph{NIPS workshop on deep learning and unsupervised feature learning}, volume 2011, page~4. Granada, 2011.

\bibitem[Nguyen and Tran(2021)]{Wanet}
A.~Nguyen and A.~Tran.
\newblock Wanet--imperceptible warping-based backdoor attack.
\newblock In \emph{International Conference on Learning Representations}, 2021.

\bibitem[Nguyen et~al.(2023)Nguyen, Nguyen, Tran, Doan, and Wong]{IBA}
T.~D. Nguyen, T.~A. Nguyen, A.~Tran, K.~D. Doan, and K.-S. Wong.
\newblock Iba: Towards irreversible backdoor attacks in federated learning.
\newblock \emph{Advances in Neural Information Processing Systems}, 36:\penalty0 66364--66376, 2023.

\bibitem[Qian et~al.(2024)Qian, Wu, Wei, Ding, Xiao, Xiang, Ma, and Guo]{EmInspector}
Y.~Qian, S.~Wu, K.~Wei, M.~Ding, D.~Xiao, T.~Xiang, C.~Ma, and S.~Guo.
\newblock Eminspector: Combating backdoor attacks in federated self-supervised learning through embedding inspection.
\newblock \emph{arXiv preprint arXiv:2405.13080}, 2024.

\bibitem[Rehman et~al.(2023)Rehman, Gao, De~Gusm{\~a}o, Alibeigi, Shen, and Lane]{L-dawa}
Y.~A.~U. Rehman, Y.~Gao, P.~P.~B. De~Gusm{\~a}o, M.~Alibeigi, J.~Shen, and N.~D. Lane.
\newblock L-dawa: Layer-wise divergence aware weight aggregation in federated self-supervised visual representation learning.
\newblock In \emph{Proceedings of the IEEE/CVF international conference on computer vision}, pages 16464--16473, 2023.

\bibitem[Ronneberger et~al.(2015)Ronneberger, Fischer, and Brox]{u-net}
O.~Ronneberger, P.~Fischer, and T.~Brox.
\newblock U-net: Convolutional networks for biomedical image segmentation.
\newblock In \emph{Medical image computing and computer-assisted intervention--MICCAI 2015: 18th international conference, Munich, Germany, October 5-9, 2015, proceedings, part III 18}, pages 234--241. Springer, 2015.

\bibitem[Russakovsky et~al.(2015)Russakovsky, Deng, Su, Krause, Satheesh, Ma, Huang, Karpathy, Khosla, Bernstein, et~al.]{imagenet}
O.~Russakovsky, J.~Deng, H.~Su, J.~Krause, S.~Satheesh, S.~Ma, Z.~Huang, A.~Karpathy, A.~Khosla, M.~Bernstein, et~al.
\newblock Imagenet large scale visual recognition challenge.
\newblock \emph{International journal of computer vision}, 115:\penalty0 211--252, 2015.

\bibitem[Saha et~al.(2022)Saha, Tejankar, Koohpayegani, and Pirsiavash]{BASSL}
A.~Saha, A.~Tejankar, S.~A. Koohpayegani, and H.~Pirsiavash.
\newblock Backdoor attacks on self-supervised learning.
\newblock In \emph{Proceedings of the IEEE/CVF Conference on Computer Vision and Pattern Recognition}, pages 13337--13346, 2022.

\bibitem[Selvaraju et~al.(2017)Selvaraju, Cogswell, Das, Vedantam, Parikh, and Batra]{Grad-cam}
R.~R. Selvaraju, M.~Cogswell, A.~Das, R.~Vedantam, D.~Parikh, and D.~Batra.
\newblock Grad-cam: Visual explanations from deep networks via gradient-based localization.
\newblock In \emph{Proceedings of the IEEE international conference on computer vision}, pages 618--626, 2017.

\bibitem[Stallkamp et~al.(2012)Stallkamp, Schlipsing, Salmen, and Igel]{GTSRB}
J.~Stallkamp, M.~Schlipsing, J.~Salmen, and C.~Igel.
\newblock Man vs. computer: Benchmarking machine learning algorithms for traffic sign recognition.
\newblock \emph{Neural networks}, 32:\penalty0 323--332, 2012.

\bibitem[Tao et~al.(2024)Tao, Wang, Feng, Shen, Ma, and Zhang]{DRUPE}
G.~Tao, Z.~Wang, S.~Feng, G.~Shen, S.~Ma, and X.~Zhang.
\newblock Distribution preserving backdoor attack in self-supervised learning.
\newblock In \emph{2024 IEEE Symposium on Security and Privacy (SP)}, pages 2029--2047. IEEE, 2024.

\bibitem[Van~Berlo et~al.(2020)Van~Berlo, Saeed, and Ozcelebi]{early2}
B.~Van~Berlo, A.~Saeed, and T.~Ozcelebi.
\newblock Towards federated unsupervised representation learning.
\newblock In \emph{Proceedings of the third ACM international workshop on edge systems, analytics and networking}, pages 31--36, 2020.

\bibitem[Wang et~al.(2019)Wang, Yao, Shan, Li, Viswanath, Zheng, and Zhao]{Neural_cleanse}
B.~Wang, Y.~Yao, S.~Shan, H.~Li, B.~Viswanath, H.~Zheng, and B.~Y. Zhao.
\newblock Neural cleanse: Identifying and mitigating backdoor attacks in neural networks.
\newblock In \emph{2019 IEEE symposium on security and privacy (SP)}, pages 707--723. IEEE, 2019.

\bibitem[Wang et~al.(2004)Wang, Bovik, Sheikh, and Simoncelli]{ssim}
Z.~Wang, A.~C. Bovik, H.~R. Sheikh, and E.~P. Simoncelli.
\newblock Image quality assessment: from error visibility to structural similarity.
\newblock \emph{IEEE transactions on image processing}, 13\penalty0 (4):\penalty0 600--612, 2004.

\bibitem[Wu et~al.(2024)Wu, Ma, Wei, Ding, Yang, and Qian]{uba}
S.~Wu, C.~Ma, K.~Wei, M.~Ding, J.~Yang, and Y.~Qian.
\newblock Towards efficient backdoor attacks against federated self-supervised learning as a service through intra-union aggregation.
\newblock In \emph{International Conference on Service Science}, pages 122--135. Springer, 2024.

\bibitem[Zhang et~al.(2023)Zhang, Kuang, Chen, You, Shen, Xiao, Zhang, Wu, Wu, Zhuang, et~al.]{FedCA}
F.~Zhang, K.~Kuang, L.~Chen, Z.~You, T.~Shen, J.~Xiao, Y.~Zhang, C.~Wu, F.~Wu, Y.~Zhuang, et~al.
\newblock Federated unsupervised representation learning.
\newblock \emph{Frontiers of Information Technology \& Electronic Engineering}, 24\penalty0 (8):\penalty0 1181--1193, 2023.

\bibitem[Zhang et~al.(2024{\natexlab{a}})Zhang, Wang, Han, Jin, Zhan, Du, Wang, and Ma]{IMPERATIVE}
H.~Zhang, Z.~Wang, T.~Han, M.~Jin, C.~Zhan, M.~Du, H.~Wang, and S.~Ma.
\newblock Towards imperceptible backdoor attack in self-supervised learning.
\newblock \emph{arXiv preprint arXiv:2405.14672}, 2024{\natexlab{a}}.

\bibitem[Zhang et~al.(2024{\natexlab{b}})Zhang, Liu, Jia, and Gong]{CorruptEncoder}
J.~Zhang, H.~Liu, J.~Jia, and N.~Z. Gong.
\newblock Data poisoning based backdoor attacks to contrastive learning.
\newblock In \emph{Proceedings of the IEEE/CVF Conference on Computer Vision and Pattern Recognition}, pages 24357--24366, 2024{\natexlab{b}}.

\bibitem[Zhang et~al.(2024{\natexlab{c}})Zhang, Zhu, Di~Wu, Yong, and Long]{BADFSS}
J.~Zhang, C.~Zhu, X.~S. Di~Wu, J.~Yong, and G.~Long.
\newblock Badfss: Backdoor attacks on federated self-supervised learning.
\newblock In \emph{Proceedings of the 33rd International Joint Conference on Artificial Intelligence (IJCAI-24)}, pages 548--558, 2024{\natexlab{c}}.

\bibitem[Zhang et~al.(2018)Zhang, Isola, Efros, Shechtman, and Wang]{lpips}
R.~Zhang, P.~Isola, A.~A. Efros, E.~Shechtman, and O.~Wang.
\newblock The unreasonable effectiveness of deep features as a perceptual metric.
\newblock In \emph{Proceedings of the IEEE conference on computer vision and pattern recognition}, pages 586--595, 2018.

\bibitem[Zhuang et~al.(2022)Zhuang, Wen, and Zhang]{FedEMA}
W.~Zhuang, Y.~Wen, and S.~Zhang.
\newblock Divergence-aware federated self-supervised learning.
\newblock In \emph{International Conference on Learning Representations}, 2022.

\end{thebibliography}
